\documentclass[12pt,epsf]{article}

\def\bOmega{\mbox{\boldmath$\Omega$}}

\def\bF{{\bf F}}

\newcommand{\be}{\begin{equation}}
\newcommand{\ee}{\end{equation}}
\newcommand{\bea}{\begin{eqnarray}}
\newcommand{\eea}{\end{eqnarray}}

\newcommand{\lan}{\langle}
\newcommand{\ran}{\rangle}

\usepackage[dvips]{graphics}

\begin{document}

\vskip 2cm
\begin{center}
\Large
{\bf Syntactic and Semantic Distribution } \\
{\bf in } \\
{\bf Quantum Measurement } \\
\vskip 0.5cm
\large
Ken Williams \\
{\small \sl Ojai, CA. } \\
\end{center}
\thispagestyle{empty}
\vskip 0.7cm

\begin{abstract}

The nondistributivity of compound quantum mechanical propositions leads to a
theorem that rules out the possibility of microscopic deterministic hidden
variables, the Logical No-Go Theorem. We observe that there appear in fact
two distinct nondistributivity relations in the derivation: one with a
semantics governed by an empirical conjunctive syntax, the other composed
of conjunctive primitives in the quantum mechanical probability calculus. We
venture to speculate how the two come to be confused in the derivation of
the theorem.
\end{abstract}

\newpage

\section{introduction}

The 20th century witnessed a revolution in experimental instrumentation from 
the
likes of the Plank's black box apparatus to the Stern-Gerlach spin analyzer.
From these there came a wealth of new and unusual data, much of which
suggested a microscopic substructure \cite{64} whose
workings were
not governed by the then prevailing Newtonian mechanics. Eventually,
particles came to be
seen no longer as entities with categorical properties but as carriers
of properties that could only be inferred from the experimental
probabilities that they collectively generate. In what
has become the orthodox interpretation of
the data and governing theory, quantum mechanics (QM), the reasoning is
taken so far as to call into
question the very notion causality implicit to scientific reasoning and as
such continues to present to the interested student an array of
counterintuitive conceptual challenges.

To elaborate the new conception there has over the years come several formalizations of the quantum theory
\footnote{ Schrodinger's wave mechanics, Heisenberg's matrix mechanics,
Dirac and von Neumann's Hilbert space formulation, Feynman's path integral
formulation, von Neumann and Segal's c-algebra formalism, Everett's many
worlds interpretation, Gell-Mann and Griffith's consistent histories
formulation, quantum logic formulation advanced by  von Neumann
and Mackey, and others \cite{54}} whose
profusion and variety however it now seems may well have had much of an
opposite effect. But a great many of these may be understood as partial
interpretations of the original consistent mathematical formulation credited to Von Neumann \cite{18} and Dirac, and whose modern version
is now standard to most QM texts; to understand this formulation then is to
understand the foundation upon which many of the others are built.

A distinctive feature of the von Neumann formal machinery is its
axiomatization of
indeterminacy as fundamental to microscopic events. This
is posited via the Collapse Postulate \cite{200}, which helps account for the
ubiquitous dispersion of ensemble experimental values on one hand \cite{65}
and for
the observation of definite individual experimental outcomes on the other.
Another distinctive feature is its representation of the experimental
process by the
action of Hilbert space operators that among themselves generally do not
commute; this to account for the observation that pairs of consecutive
measurements performed on a single system when temporally reversed generally
do not yield the same pair of outcomes, i.e., for the observation that such
measurements also do not commute. Around these core ideas has developed an
increasingly abstract semantics - rules that lay down the correspondence
between the theoretical terms in the mathematical machinery of quantum
mechanics and observation - now a source of conceptual difficulty for and
disagreement among all interested parties from physicist to philosopher.
E.g., what to a Bayesian inclined mathematician or philosopher are
relations concerning the uncertainty of individual experimental outcomes
\cite{1}, to an empirically minded experimentalist may be nothing more
than unusual scatter relations \cite{10}. And so forth.

This particular example highlights the central question of concern to an
interpretation of the theory: Whether it is
possible to
supplement the quantum mechanical description of reality with additional
parameters, so called hidden variables (hv), which would then together give
a more 'complete'
account
of microscopic processes and states, including absent in the existing
theory, such as those states that correspond to
noncommuting experimental outcomes. On this issue there is certainly a wide
range of possible views, but the
leading majority opinions, as a matter of fact and history, are and have
been polarized. In
the affirmative view, whose early proponents include A. Einstein, the
proposed
notion of the quantum particle is at odds with the very concept of
'particle'
conceived classically as a point in phase space, and the incompleteness of
the theory is self-evident. Those in opposition, proponents of
the
conventional or orthodox interpretation, have gone so far as to produce
explicit
proofs against the very possibility, somehow managing to prove a negative.

Among these proof, popularly known as'no-go' theorems, perhaps the best known is the one due
exclusively to John Bell. By exploiting the locality requirements of special
relativity Bell derives an explicit disagreement between the tenants of the
local realism and the predictions of QM, summarized
in his elegant inequality \cite{11}. Next in order of the interest it has
generated over the years is the theorem of Kochen and Specker (KS) who begin by taking
the possibility of isomorphisms from the Hilbert subspaces of QM
to classical Boolean subspaces as a basic constraint on realist interpretations, then demonstrates that there are none. The
significance of each of these is addressed by the writer in earlier works
\cite{19, 20}. Finally there is the lesser known argument against hidden
variables advanced in the mid-sixties by Jauch and Piron, the
Logical no-go theorem \cite{12}. Interest in this proof however peaked and
quickly
declined until it is today not much discussed at all \footnote{Thompson's
ISI Web of Knowledge lists 1934 citations of Bell's theorem, 381 of the
theorem of KS, and 92 of the logical no-go theorem, only 10 of those since
2000.}, the
remaining interest lying mainly in its close association with an earlier
and
similar argument by von Neumann and with the later work of Kochen and
Specker. More importantly still is its place in the historical development
of logical formulations of the theory, of quantum logics.

To briefly outline the basic quantum logic idea, to every experimental
outcome there
corresponds a proposition (for outcome 'a', the proposition: 'the
experimental outcome is a'). Then the indeterminacy of measurement outcomes
as axiomatized by von Neumann imposes in an obvious way a certain
non-bivalence upon the truth values of the individual propositions
corresponding to those outcomes (such that all experimental propositions in
respect of the physical system upon which measurements are to be taken,
experiments performed, are not of necessity either true or false \cite{66}), i.e.
upon the
truth values of individual propositions, their system then
corresponding to the set of all measurements that may be made upon the given
physical system. Thus, structural features inherent to the standard
formulation's Hilbert space, whose operators are bijective \cite{100} in
respect of possible
observations, correspond directly to those of the proposition system. By
means of semantical rules, these in turn correspond, presumably, to logical
structures
extant in the microscopic physical world and so now framed in the language
of
quantum mechanics. But such a system and a logic, like their Hilbert space
description, are
non-Boolean, hence non-classical.

Those familiar with quantum theory will probably have first encountered this distinction in some form or other of Bohr's complementarity \cite{110},
as complementary variables are also variables that do not commute; hence,
the logic of their observation or measurement is necessarily non-Boolean.
Given the recent important experimental {\it welcher-weg} tests conducted by S. Afshar and students
\cite{111} and the questions concerning
complementarity raised by their results, still under review \footnote{
For a rebuttal see W. Unruh's article at URL = $ \lan $
http://axion.physics.ubc.ca/rebel.html  $ \ran $.  }, a critical review of the complementary semantics, such as
the present one, that also maintains an elementary presentation, could
hardly seem to us more timely.

In this article we analyze the particular Logical argument against the
existence of hv's put forward by Jauch and Piron \cite{12},
whose driving force we trace to a semantical rule for the
conjunction of
propositions, $ a \bigcap b $, associated with pairs of measurements that
do not commute, $ [a, b] \neq 0 $. While the set theoretic and ordinary
logical semantics of the conjunction are well known, the compound being true
when each proposition is separately true, in the new logic there remain
questions. The syntactic structure has been analyzed over the years by many
workers in the field, and in the view of some
\cite{2,12,13,14,18,26}, prima facie in line with its Hilbert space
correspondence, as an experimental proposition the noncommutative
conjunction is tautologically false. However according to others
\cite{3,6,16,17,42}, and in line with more direct semantics, such
compounds are not
experimental propositions that bear on {\it individual systems} at all,
but
are in this respect formal expressions having no real meaning. It is
entirely possible
that this issue cannot be settled objectively, as the difference in
opinion may be grounded in the much longer standing difference in
interpretation of
the quantum theory itself. In this article we attempt to understand the
noncommutative conjunction exclusively in terms of its use.

In section 1 we consider a microscopic experiment instrumental in motivating
the conceptual development of quantum theory and trace the noncommutative
conjunctive etymology within its logic to the sigma algebra of  its Hilbert $ H
\bigcup H^\prime $ probability space whose nondistributive syntax reveals
the dispersive semantics that leads to the conclusions of the logical no-go
theorem:
\begin{center}
realist interpretation $\Rightarrow $ value-definiteness $\Rightarrow $
dispersion-free mixtures $\Rightarrow $ distributive logic $\Rightarrow $
commutative logic $\Rightarrow $ classical physics.
\end{center}
We follow up in the next section with an examination of the probability
space of
compound noncommuting observations where we find a formally identical
nondistributivity relation which, in contrast to the previous relation, is
grounded in the metalanguage of the theory. There, the noncommuting
'conjunction' appears as an elementary or atomic event in the product $ H
\times H^\prime $ space \cite{101}. In light of the apparent distinct events
they reference, we consider in the next section whether the two relations
might in
fact correspond to the same physical property. We find that they do not (at
least from the relevant realist point of view), which then invalidates the
logical no-go theorem. And while both Bell and Bohn challenged this validity long ago,
their results were generally not well received
at the time.

In addition to our adherence to an elementary exposition remaining within
the purview of
undergraduate QM, another difference between the earlier analysis and our
approach is the
region of analytic validity that we concede to the theorem; we observe that
the opposing views operate on distinct semantics that follow, in one case,
from the
syntactic reduction of diatomic compound bivalent experimental
propositions, and in the other, from that of such propositions over their
aggregates
(which are generally nonbivalent), then combined. This point of
view offers we think a more comprehensive understanding of the disagreement.
We end in the final section with a few concluding remarks on the theorem and
related issues.

While the concepts central to the logical no-go theorem and quantum logic generally 
are fundamentally simple, they do involve a myriad of definitions and
notations
unfamiliar to most students and non-specialists, although again, no single one
of these particularly difficult to grasp. It is also likely that many
readers will
first encounter this article via an internet resource. For these reasons we
make extensive use of internet citations. We often point to Wolfram's {\it
MathWorld and the Statistics Glossary} for clarifications and basic
definitions in probability theory, and  to the {\it The Stanford
Encyclopedia of Philosophy}, {\it  The Philosophy Pages}, and {\it
Wikipedia, the free encyclopedia} for philosophical and historical
contexts.

\section{ argument against hidden variables}

With an aim to predict and finally manipulate physical events
and
processes, the scientific enterprise proceeds on the implicit premise that
given the relevant physical laws and prevailing conditions, the occurrence of
subsequent events may in principle always be known beforehand. I.e., it
proceeds on the premise that such physical laws indeed exist and is thus
fundamentally entrenched in the determinism hypothesis. It is an irony then
that
the facilitating scientific method, famously successful in
hypothesis self-correction, is itself not subject to the same
correction, as there is no rule to tell us just when the
determinism hypothesis
breaks down, no negative test of the hypothesis. The rule of practice, as
part and parcel of the method itself, is that the hypothesis never does
breaks
down; it is the unsatisfactory prediction itself that motivates the
search for causation, Newton pondering the fallen apple. The final
justification of the method rests, as always, in the likelihood of future
discovery.

It is in the event of unknown and thus possibly nonexistent physical laws
that the program may run afoul the prevailing "belief that natural
science, based on observation, comprises the whole of human knowledge", to
quote from the Philosophy Pages  entry for Positivism \cite{24},  where in
the extreme view further elaborated, whatever the rational appeal or past
successes
of the determinism hypothesis, non-empirical statements of all brands
are metaphysical \cite{25}. Upon this reasoning an epistimically
undetermined microscopic experimental outcome becomes, in accordance with von
Neumann's reduction axiom and corroborating Copenhagen interpretation of
QM, ontologically indeterminate. But more on this later. It
should at least be clear that a probabilistic theory understood also as
complete (in respect of its account of the physically objective world) such
that empirical collective statistics at the same time
exhaustively characterize each collective-member also  - such a theory
naturally assumes a strongly subjective (e.g. Bayesian) quality.

In our lead-up to the logical no-go derivation ( whose standard presentation
is couched
in a specialized nomenclature), we first, in the next section introduce the
necessary experimental
and formal terminology by way of considering an application of the QM
probability theory to a specific instance.

\subsection{the structure of experimental outcomes}

Let us consider a physical system and the set of experiments that may be
performed on it. To each experiment there corresponds an array of
characteristic outcomes, an experimental spectrum, which for a sufficiently
large number of identical experimental trials may then be mapped to a
probability
distribution, a state space, each element of which being equal to the
long-run relative frequency recorded for the corresponding outcome. We
consider the complete set of such distinct experimental processes. To the compound
mapping then there corresponds a parallel mapping from experimental
propositions (as we have seen, corresponding to outcome 'a', the
proposition: 'the outcome
is a') to the interval [0,1], taken as a measure of the truth of a given
proposition: mapped to '1' for true, to '0' for not-true. And like its
experimental counterpart, this mapping too is generally non-injective
\cite{63}, as
distinct experimental arrangements may sometimes yield identical results;
i.e., outcomes sometimes overlap (Classically, e.g., the 'weight' measurement outcome for a
given mass on
earth will be identical to the weight measurement, say on the moon, of an entirely different mass.).

The sort of quantum experimental data that readily lends itself to this description
is obtained from measurements of microscopic spin of the kind taken in
Stern-Gerlach (SG) experiments  \cite{52,56}. There, an assemblage, or
ensemble, of identically prepared particles is accelerated through a
localized inhomogeneous magnetic field from which they emerge with
velocities in one of a discrete number of directions
\begin{center}
\scalebox{0.7}[0.7]{\includegraphics{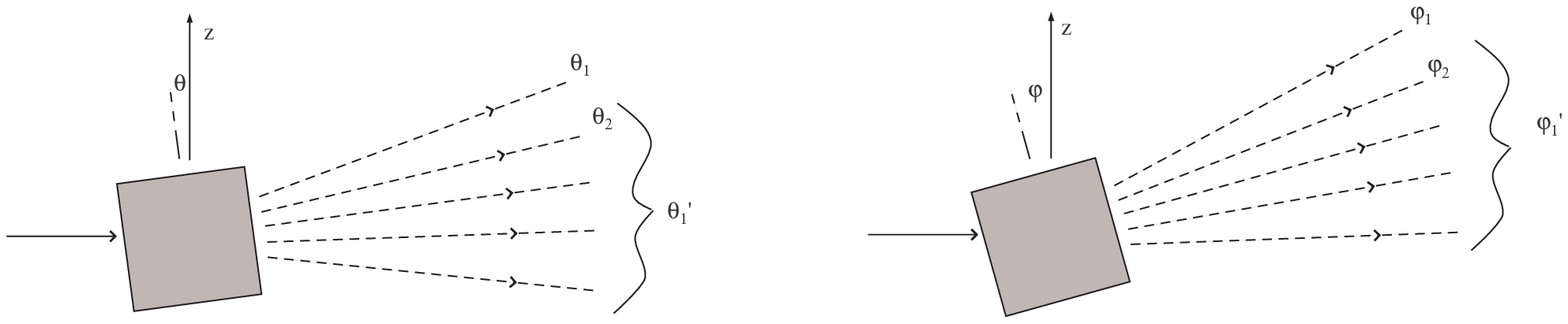}}\\
figure 1
\end{center}
a given direction characteristic of a particle's spin projection along the
SG symmetry
axis. We know however from experience with ordinary macroscopic spins and
from the predictions of classical electromagnetism that these directions
should instead vary continuously
\begin{center}
\scalebox{1.0}[1.0]{\includegraphics{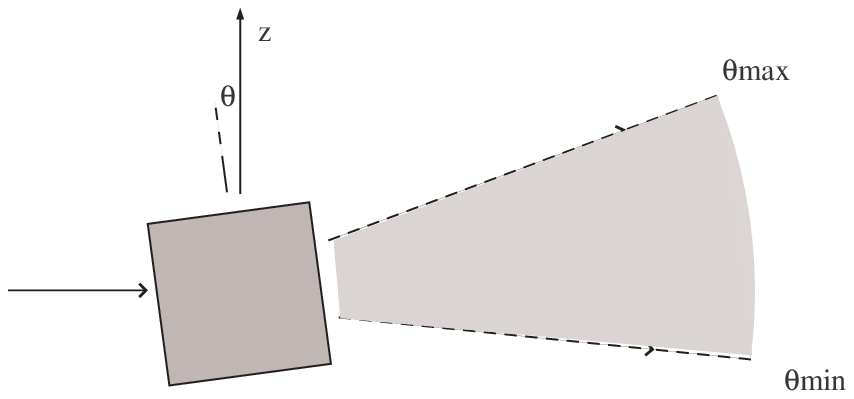}}\\
figure 2
\end{center}
with limits determined by the spin magnitude and SG field strength.
Microscopic spins predictions are for this reason said to be 'quantized',
appearing, observed, only in discrete amounts, and a SG experiment $ \theta
$ is thus
characterized by its {\it discrete} outcome set $ \Omega_\theta = \{
\theta_1, \theta_2, \theta_3, .... \theta_n \} $, with outcome probabilities
given by the experimental relative frequencies
\bea
{ \rm P }( \theta_i ) & = & { \rm p}_i  =  n_i /n  \nonumber
\eea
\bea
\sum n_i & = & n, \quad { \rm so \, \, that } , \quad \sum {\rm p }_i = 1
\nonumber
\eea
where $ n_i $ is the number of experimental trials with outcome $
\theta_i
$ , and n the total number of trials in a given run  \cite{52}.

In respect of the formal probability space $ ( \Omega_\theta, { \rm \bF,  P
} )  $, the probability measure P maps \bF \, to the reals,
P: \bF $ \rightarrow $  [1,0], where $ \bF $ is the  sigma algebra generated
by $ \Omega_\theta $, and is thus composed of the closed unions of subsets
of $ \Omega_\theta , \, E_i \bigcup E_j  $, called events,  where $ E
\subseteq \Omega  $ . Events then are sigma-measurable subsets and may
always be expanded as a finite union of outcomes
\bea
{\rm E } &=& \{ \theta_i \} \bigcup \{ \theta_j \} \bigcup \{ \theta_k \}
\ldots  = \{ \theta_i , \theta_j, \theta_k, \ldots \} \nonumber
\eea
for which expansion we use the notation
\bea
{\rm E } &=&  \theta_i  \bigcup  \theta_j  \bigcup  \theta_k  \ldots
\eea
Elements of \bF  are called the measurable or Borel sets pertaining to the
given experiment, while the probability measure P has the property $ {\rm P
}( {\rm E }_i \bigcup {\rm E }_j ) =  {\rm P }( {\rm E }_i ) + {\rm P }(
{\rm E }_j ) $
whenever events $E_i $ and $ E_j $ are disjoint, denoted, $ {\rm E }_i \perp
{\rm E }_j $  \cite{38}. Then with
\bea
{ \bf F } &=& \{ \theta_1, \theta_2, \theta_3, \ldots , \theta_n , \theta_1
\bigcup \theta_2,  \theta_1 \bigcup \theta_3, \ldots ,  \theta_i \bigcup
\theta_j \bigcup  \theta_k, \ldots \ldots ,   \theta_1 \bigcup \theta_2
\bigcup \ldots \bigcup \theta_n  \} \nonumber
\eea
single element events, here $\theta_1 $, $\theta_2$, and $ \theta_3 $, represent individual experimental outcomes and are
said to be atomic or elementary; they are the {\it primitive} elements
of the probability theory, external inputs of a truth value status
independent of the
theory, while
the general $ \bF $ element represents combinations of individual outcomes.
As an experimental probability mapping is characteristic of the
corresponding ensemble of observations, distinct formal probability
functions P may be taken to represent distinct states of the ensemble.

It is possible to generalize the outcome set by taking at once the union of
all outcome sets, $ \Omega_\theta  \rightarrow   \Omega_\theta  \bigcup
\Omega_\varphi \bigcup   \Omega_\chi \bigcup \ldots =  { \rm X }  $, called
the outcome space \cite{2} \footnote{One might well question whether the
criterion of "generalization" is met here. More on this later}.
\bea
\Omega \rightarrow {\rm X } &=& \{ \theta_1, \theta_2, \theta_3, \ldots,
\theta_n ,  \varphi_1, \varphi_2, \varphi_3, \ldots , \varphi_n ,  \chi_1,
\chi_2, \chi_3, \ldots , \chi_n , \ldots \ldots  \}        \label{space}
\eea
A peculiarity of
measurements on microscopic ensembles is the absence of experimental
mappings P: X $ \rightarrow $  \{1,0 \} such that all ensemble members have,
simultaneously, all the same projections. The phenomena is called
dispersion; thus, all microscopic ensembles are observed to be dispersive
\cite{65}.

\subsection{ formal structures in a Hilbert space}

It happens that the forgoing formal relations are structured in a manner
similar to those among the elements in a vector space. We consider then an
n-dimensional Hilbert space (H-space) spanned by the representative basis
\bea
\bOmega &=& \{  | 1 \ran ,  | 2 \ran ,   | 3 \ran ,  \ldots , | n \ran    \}
\nonumber
\eea
The span of this basis, comprised of all possible linear combinations, $
\alpha_1 | 1 \ran +   \alpha_2 | 2 \ran +   \alpha_3 | 3 \ran +  \ldots +
\alpha_n | n \ran $, where the $ \alpha _i $  are complex numbers, constitutes the H-space itself. Among the basis elements are the structural
relations  \cite{53}
\bea
\lan i | j \ran &=& \delta_{ij}  \quad \quad \quad \quad { \rm  (
orthonormalization ) }      \nonumber \\
\sum | i \ran \lan i | &=& { \cal    I }_{n \times n } \quad \quad \quad
\quad  {\rm  ( completeness ) } \nonumber
\eea
by means of which one orthonormal basis is related to another:  $ |j^\prime
\ran = {\bf I }_{n \times n } | j^\prime \ran  = ( \sum | i \ran \lan i |
) | j^\prime \ran = \sum \lan i | j^\prime \ran | i \ran = \sum c_{j^\prime
i } | i \ran $.
To each unit element then there corresponds a characteristic
operator that projects any H-space vector $ | \psi \ran  $  onto and so
defines a unique subspace, $ ( | k \ran \lan k | ) | \psi \ran = c_{k \psi }
|k \ran  $,  for some $  c_{k \psi }  < 1 $. The operator $ P_k =  | k
\ran \lan k |  $, thus projects an arbitrary vector onto the $ H_k $
subspace $ \{ | k \ran \}  $, and is known as a {\it projection operator}
\bea
P_k | \psi \ran & =& c_{k \psi } |k \ran  \, .       \nonumber
\eea
The complete H-space is then a formal union of such subspaces
\bea
{\rm H } &=&  {\rm H }_1 \bigcup {\rm H }_1 \bigcup{\rm H }_2 \bigcup{\rm H
}_3 \bigcup \ldots  {\rm H }_n      \nonumber \\
&=& {\rm H }_i \bigcup{\rm H }^\prime_i \nonumber
\eea
where $ { \rm H}_i^\prime $  here is the H space relative complement
\cite{31} to the $ {
\rm H}_i $ subspace.

\subsection{ semantical rules}

When we now assign an experimental outcome set to an orthonormal basis, $
\Omega \sim \bOmega  $, and thus $ X \sim H $, we identify a pre-measured
state such as $ \psi $  in figure1 with an expanded vector in this basis
\bea
\psi  & \sim& | \psi \ran = \sum c_{i \psi} | i \ran \\       \nonumber
&& \sum c_{i \psi}^2 =1 \nonumber
\eea
and obtain the experimental statistics, the observed distribution, P:  $
\Omega \rightarrow \{ {\rm p }_i \}  $, by means of the scalar product $
\lan \psi | j \ran   $  as
\bea
{\rm p}_i  &=& | \lan \psi | i \ran    |^2   \, .      \nonumber
\eea
Further, the observed ensemble dispersion manifests here as the nonexistence
of
H-space vectors $ | \phi \ran $  having the property
\bea
\lan n^\prime | \phi \ran &=& 0 \, {\rm or } \, 1 , \quad \quad { \rm  for
\, \, all   } \,  n^\prime     \nonumber
\eea
In other words, there can be no probability measure, no state, with
the property, $ {\rm P }_\psi : H \rightarrow \{ 0, 1 \} $.

In terms of projectors, the previous H-space structure relations become
\bea
P_i P_j &=& \delta_{ij} P_j  \quad \quad \quad { \rm  ( orthonormalization
)  }      \nonumber \\
\sum P_i &=& {\bf I }_{ n \times n } \quad \quad \quad { \rm
(completeness)   } \nonumber \\
{\rm p }_i &=& | \lan \psi | P_i | \psi \ran  |   \nonumber
\eea
The main advantage of this formulation lies in the correspondence between
projection operators and experimental propositions.
The projectors are QM operators with eigenvalue set \{0,1\}, so that as a
projector corresponds to an experimental proposition, $ ( \theta_i \sim | i
\ran \sim P_i ) $,  its eigenvalue corresponds to the proposition's truth
value: '1' for 'true', '0' for 'false'; likewise, as the probability $ p_i $
  gives the projection of $ | \psi \ran $ along $ | i \ran $, the
corresponding projector maps to the proposition $ \theta_i $ , the
proposition that $  P_\psi \subseteq P_i (  H_\psi \subseteq H_i ) $. As a
consequence, in this vector-{\it space} formulation of states we have that
\bea
P_i & \subseteq & {\bf I }_{j } ( H_i \subseteq H_{j} \bigcup H^\prime_{j }
) , \quad \quad {\rm for \, \, all } \,\, \theta_i \, {\rm and } \,\,
\varphi_j
         \nonumber
\eea
whereas in a vector-{\it set} formulation we have, as in set theory, $
\theta_i \subseteq (
\varphi_j \bigcup \varphi_j^\prime  )  $, only in the event that either $
\theta_i  \subseteq  \varphi_j \, \, {\rm or } \,
\, \theta_i  \subseteq  \varphi_j^\prime  $ .

\subsection{logical structure of micro-events}

By the logical structure of microscopic events we refer the interrelations
among the propositions that assert the occurrence of such events. And as to
each individual experimental outcome there is assigned  a yes-no probability
distribution, to the corresponding proposition is assigned a truth-value
distribution, the two distributions, presumably, being one and the same.

Among experimental propositions, and propositions in general, there are
ordering relations of implication, such that the truth of one proposition
may imply that of another. This relation is typically expressed in the
notation of naive set theory as set inclusion, $ \theta_i \subseteq
\varphi_j $, here $ \theta_i $  implying $ \varphi_j  $. Whereas an
equivalence of propositions, $ \theta_i = \varphi_j $, simply represents
the
combined orderings $ \theta_i \subseteq \varphi_j $ and $ \varphi_j
\subseteq \theta_i $. Consider for example a case in which two volumes
physically overlap, $ V_a \subseteq V_b $, and the proposition a (b): the
particle is in volume  $ V_{ a (b )} $. It is then by self-evident tautology
that, a $ \subseteq  $ b, and the relation is said to be analytic. On
the other hand, there are many relations among propositions, also empirical,
such
as may embody e.g. the observation of a physical regularity or law and do
not involve tautology. For example, given propositions a: the object is
released from a height h, and b: the object reaches the ground in $ t_h $
seconds, an ordering, b $ \subseteq $  a, might express an instance of
Newton's law of gravity. Such relations as these are synthetic  \cite{50}. In both
cases, a is said to be a lower bound of b in the ordering a $
\subseteq $ b. In the set theoretic notation, the conjunction and
intersection of propositions a $ \bigcup $ b and a $   \bigcap $ b, are then
taken to be greatest lower bound (glb) and lowest upper bound (lub) of 'a or
b' and 'a and b', respectively, and are said to be true whenever 'a is true
or b is true' and 'a is true and b is true'.

The complete set of propositions bearing on the experiments that may be
performed on a given physical system constitute a proposition system
with structural properties characteristic, presumably, of the physical
system itself. As it happens, the truth structure of the conjunction of two
propositions is all important to a derivation of the logical no-go theorem.
In most analysis, the conjunction of {\it any} two
experimental propositions is again an experimental proposition having a
truth structure given by the following rule:
\begin{quote}
\it

Let I be an index set and $ \{a_i \} ( i \in I  ) $ any subset of $ L, a_i
\in L $. Then there exists a  proposition, denoted by
$ \bigcap_I  a_i $ with the property \\

\begin{center}

$ x \subseteq a_i $  for all  $ i \in I  \leftrightarrow  x \subseteq
\bigcap_I a_i $

\end{center}

\end{quote}
"axiom II" as it appears in Jauch's QM text \cite{ 13}; the proposition
system that satisfies this rule is then shown to have the structure of a
mathematical lattice  \cite{ 12,13,14,26} with
\begin{quote}
\it
\begin{center}

$ a \subseteq a $   for all   $ a \in  L $ ;  \\
$ a \subseteq b $    and  $   b \subseteq a  $   implies  $ a = b $ ; \\
$ a \subseteq b $   and $  b \subseteq c  $   implies $  a \subseteq c. $

\end{center}

To every $ a \in L $ there exists another proposition $ a^\prime \in  L $
with

\begin{center}

$ (a^\prime )^\prime = a $ ; \\
$ a^\prime \bigcap a = \oslash $ ; \\
$ a \subseteq b  \leftrightarrow  b^\prime  \subseteq a^\prime $.

\end{center}
\end{quote}
The axiom is assumed valid for experimental propositions in respect of both ordinary macroscopic and
microscopic systems \cite{64}. What sets one type apart from the
other are
the ordering relations between  propositions that bear on experiments that
do not commute, compound measurements for which the temporal order of
component application has an effect on the eventual component outcomes. For
example, if on a single physical system we perform the experimental sequence, $ \theta \varphi \theta $, resulting in $ \theta $
outcomes $ \theta_i $  and $ \theta_j $  that are not equal, i $ \neq $ j,
then experiments $ \theta $ and $ \varphi $, and corresponding propositions,
do not commute and are said to be incompatible. While the noncommutivity of
measurements on microscopic systems is readily observed, the term
'classical', sometimes ascribed to macroscopic systems, refer properly,
rather, very specifically to measurements, experiments, that commute:
classical system $ \sim $ commutative measurements on system.

\subsection{ classical versus quantum logical structures}

The truth structure of classical syntactically compound experimental
propositions is given by implicit set theoretic rules such as the law of
distribution
\bea
\theta_i \bigcap ( \theta_j \bigcup \theta_j^\prime ) &=&  ( \theta_i
\bigcap  \theta_j  ) \bigcup  (    \theta_i \bigcap  \theta_j^\prime )
\label{dis}
\eea
nicely illustrated by means of Venn diagrams
\begin{center}
\scalebox{0.7}[0.7]{\includegraphics{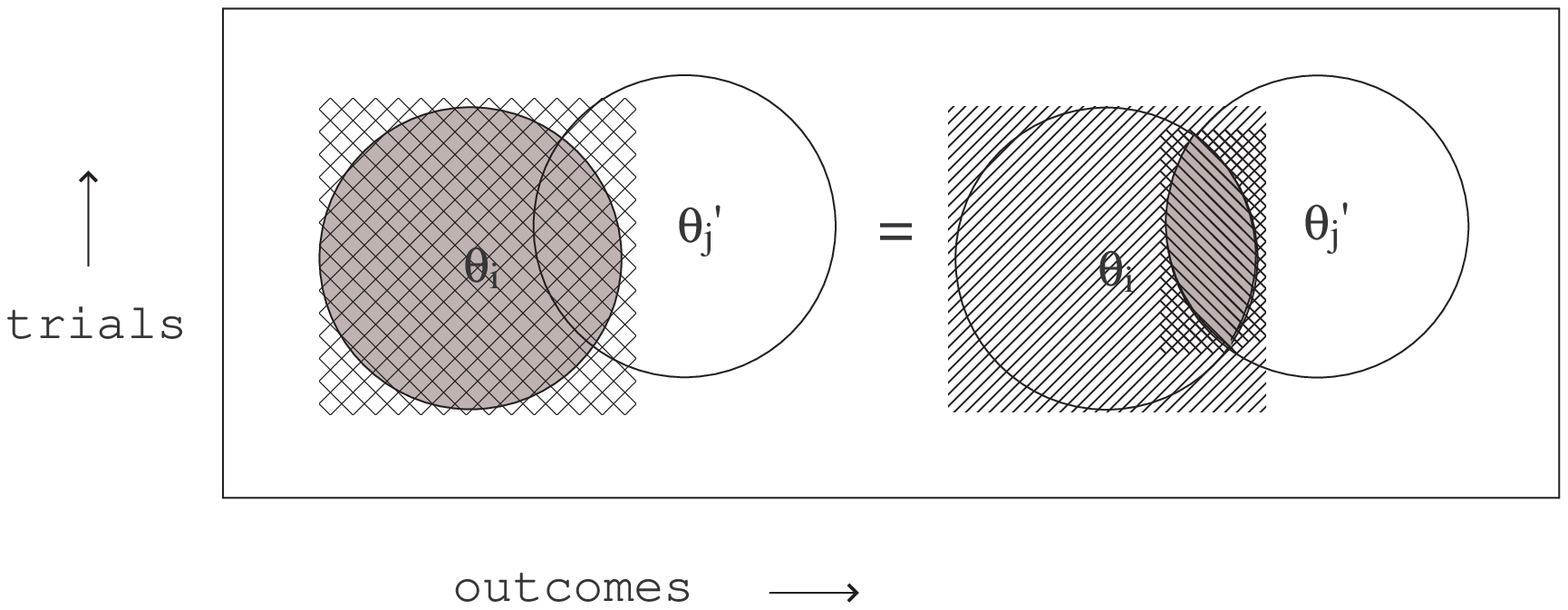}}\\
figure 3
\end{center}
where the shaded spatial areas are correlated to set size, hence to
probability.
A system of propositions obeying relation (\ref{dis}) is said to be Boolean
(or
classical)  \cite{45}.

It was the mathematician and pioneering quantum theorist Von Neumann  who
long ago first observed that mutually noncommuting propositions generally do
not satisfy the relation \cite{5}. Thus, propositional systems that refer to
classical phenomena are distributive, while those that refer to microscopic
phenomena are nondistributive. Given the significance of this distinction,
it is worth taking a close look at Von Neumann's argument as it appears in
his {\it The Logic of Quantum Mechanics} \cite{5}:  "...These facts
suggest that the distributive law {\it may} break down in quantum
mechanics. That it {\it does} break down is shown by the fact that if $ a $
denotes the experimental observation of a wave-packet $ \varphi $ on one
side
of a plane in ordinary space, $ \acute{a} $ correspondingly the observation of $
\varphi $ on the other side, and $ b $ the observation of  $ \varphi $ in a
state symmetric about the plane, then (as one can readily check)"
\begin{center}
\scalebox{0.7}[0.7]{\includegraphics{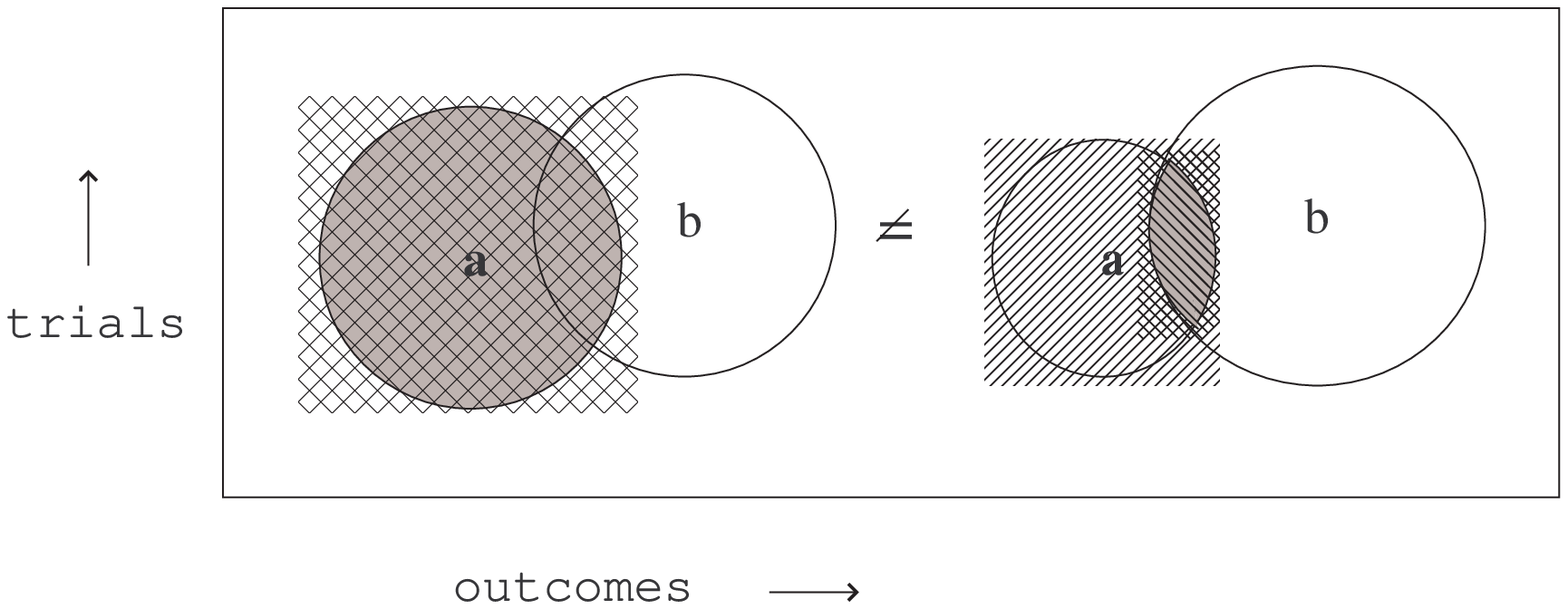}}\\
figure 4
\end{center}
\bea
a &=& a \bigcap ( b\bigcup b^\prime ) \neq (a \bigcap b ) \bigcup (a \bigcap
b^\prime )           \label{nd1}
\eea
since, actually,  (a $ \bigcap $ b) = 0 when $ a $ and $ b $ are non-collinear.
Another version of the argument in terms of spin measurement appears in
Jammer's {\it Philosophy of Quantum Mechanics} \cite{8}. From this we see
that while the logical matrix of microscopic phenomena affirms the law of
the excluded middle according to which the proposition, $ { \cal
I } $ = a $
\bigcup $ a' (a or not-a ), the proposition of 'identity', is always true,
it
denies the law of bivalence by virte of which exactly one of the
propositions 'a' or 'not-a' is of necessity true \cite{67}.

\subsection{ logical no-go theorem}

With the necessary machinery now in place it is here that we encounter
a possible conflict with the notion common to realist thinking
that to experimental processes there are causes that determine their
outcomes with certainty \footnote{By
realism we mean simply the realism e.g. characteristic of the EPR elements
of reality \cite{60,8} which presupposes determinism as a sufficient
condition \cite{19}, characteristic also of observables in Bell's theorem};
here, the beginnings to the logical
no-go. By means of this determinism experimental outcomes may in principle
always be known prior to measurement, so that future tensed propositions are
at all times bivalent,  either true or false; hence, the realist principle
of value-definiteness  \cite{ 8, 20, 27}. But the set of all microscopic
ensemble measurements, as we have seen, is empirically dispersive
\bea
\sigma (a) &\equiv& {\rm P}(a) - {\rm P}^2(a)  \neq  0 \quad \quad \quad { \rm for \,\, at
\,\, least \,\, some \,\, propositions \,\, } a   \nonumber
\eea
which then casts the realist ensemble as an assemblage of similarly prepared
though non-identical entities, as a 'mixture' of dispersion-free
sub-ensembles whose measurement yet yield the necessary (observed)
noncommutivity of
incompatible observables. Let us point out that this realist view contrasts the
previously given 'conventional' view where ensemble dispersion appears
rather as a direct manifestation of non-bivalent experimental values
possessed not only by the assemblage, but by its individual members also.
In any event, realist dispersive
ensemble states, $ \omega $, are then linear sums of
nondispersive sub-ensemble states, $ \omega_i $ .
\bea
\omega &=& \sum \alpha_i \omega_i  \quad \quad \quad \quad {\rm for \,\,
some \,\, complex \,\, numbers \,\,    } \alpha_i      \nonumber
\eea
with, for each $ \omega_i $
\bea
\omega_i (a) - \omega_i^2 (a) &=& 0 \quad \quad \quad {\rm for  \,\, all
\,\, propositions \,\, } a    \nonumber
\eea
together forming a convex set  \cite{2,13,28}. From this constraint on
realist
nondispersive subensembles, $ \omega_i $(a) = 1 or 0, for all propositions
a, it is easy to show by direct substitution into (\ref{dis}) that every
definite truth-value combination corresponding to the set of definite values
possessed by a given subensemble affirms the law of distribution. I.e.,
given propositions a and b and every possible definite value assignment,
a, b $ \in $ \{0,1\}, we find for each case
\bea
a \bigcap ( b\bigcup b^\prime ) & = &  (a \bigcap b ) \bigcup (a \bigcap
b^\prime )        \nonumber
\eea
The relation holds, recall, only in the case that measurements corresponding
to constituent propositions commute. Given that microscopic measurements generally
do not commute the realist hv interpretation of QM and its
description of the microscopic data is placed at direct odds with observation, thus
concluding the logical no-go proof.
\begin{center}
realist interpretation $\Rightarrow $ value-definiteness $\Rightarrow $
dispersion-free mixtures $\Rightarrow $ distributive logic $\Rightarrow $
commutative $\Rightarrow $ classical.
\end{center}
In the words of its authors, "Loosely stated, the main result [of the
theorem] is simply this: if there exist incompatible observables then hidden
variables are not possible." Notwithstanding critical appraisals from Bohm
and later Bell (and sadly, few others), which we discuss in a later section, at
the
time of publication this pronouncement on hidden variables enjoyed unanimous approval.

\section{ physical and logical conjunctions }

The inadmissibility of realist hv's on the grounds of
non-classical commutivity is deduced it might have been noticed from no more
than a glancing account of the phenomena of noncommutivity itself; our fig
4, for instance, which illustrates the argument connecting
noncommutation with nondistributivity advanced by von Neumann depicts no
explicit noncommutation.

\subsection{joint probabilities}

The noncommutivity of measurements is
a phenomena that necessarily involves the application of consecutive
experiments {\it
on individual physical systems}. An instance of this we denote by the
inequation, $ \varphi_j \theta_i \neq \theta_i \varphi_j $,  where the
spatial order of propositions here, right to left, represents the temporal
order of experiment application. In words, the proposition that the compound
experimental outcome is $ a $ then $ b $ generally has a different truth value
from the proposition that the outcome is $ b $ then $ a $. For comparison
with fig 1 we illustrate an experimental arrangement for which
noncommutivity effects are common.
\begin{center}
\scalebox{0.8}[0.8]{\includegraphics{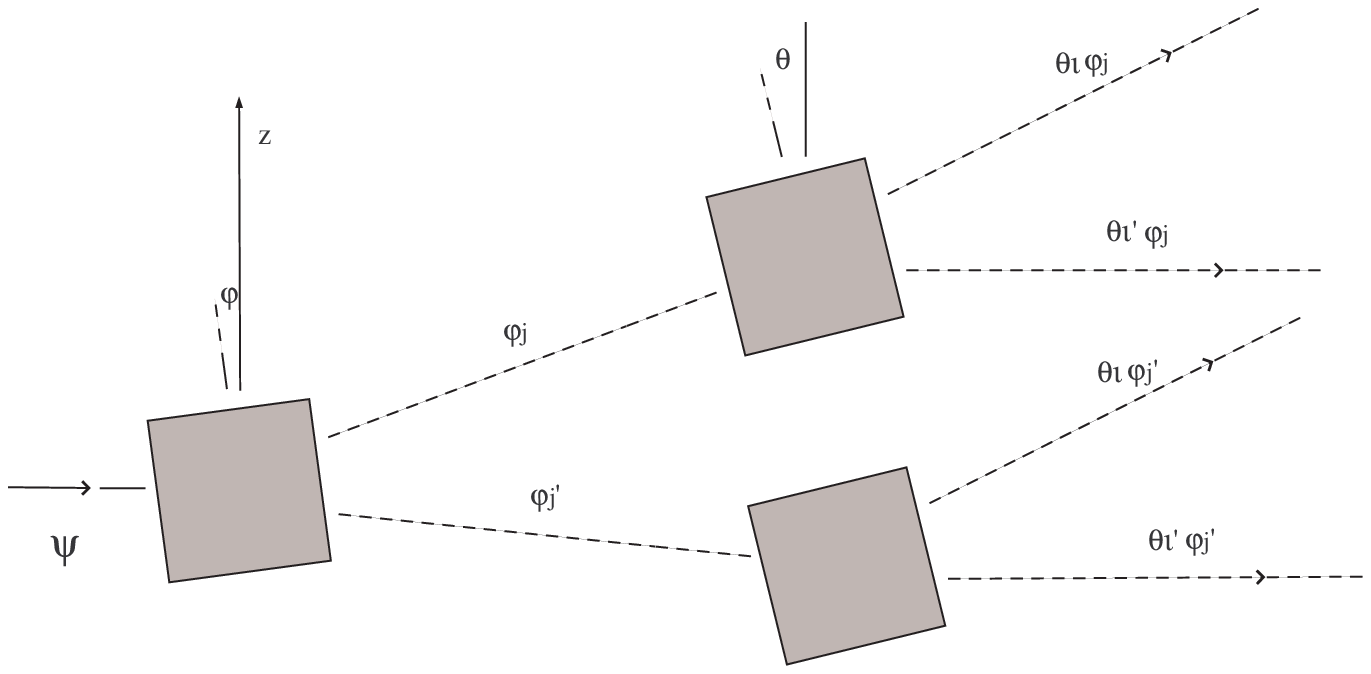}}\\
figure 5
\end{center}
Here the experimental outcome space may be given as the complete set of
possible single SG measurement outcome combinations
$  \{  \theta_i \varphi_j, \theta_i \varphi_j^\prime, \theta_i^\prime
\varphi_j, \theta_i^\prime \varphi_j^\prime\} $ ( or as the set  $ \{
\varphi_j, \varphi_j^\prime   \}  $  in terms of marginal outcomes ) whose
probabilities contribute to the marginal probability
\bea
{\rm P } ( \varphi_j ) &=&  {\rm P } (  \theta_i \varphi_j  )  +  {\rm P }(
\theta_i^\prime \varphi_j )  =  {\rm P } ( \varphi_j  | \theta_i  ) /{\rm P
} ( \theta_i ) + {\rm P }(\varphi_j  |  \theta_i^\prime)  / {\rm P }(
\theta_i^\prime )       \nonumber
\eea
The above expression P($\alpha | \beta $) here represents a conditional
probability: the truth value of proposition $ \alpha $ on the condition that
proposition $ \beta $ {\it with regard to the same physical system} has
truth value 1 \cite{41}. An exchange of labels in figure 5, $ \theta
\leftrightarrow \varphi $, obtains the converse marginal probability
\bea
{\rm P } ( \theta_i ) &=&  {\rm P } (  \varphi_j  \theta_i )  +  {\rm P }(
\varphi_j^\prime \theta_i )  =  {\rm P } ( \theta_i  | \varphi_j  )/ {\rm P
} ( \varphi_j ) + {\rm P }(\theta_i  |  \varphi_j^\prime)  /{\rm P }(
\varphi_j^\prime ) .          \label{marg}
\eea
With these definitions quantum
mechanics predicts and experiment confirms that in general
\bea
{\rm P } ( \varphi_j  | \theta_i  ) / {\rm P } ( \theta_i )  & \neq &   {\rm
P } ( \theta_i  | \varphi_j  ) / {\rm P } ( \varphi_j )   \, . \label{j1}
\eea
And yet from the same probability calculus we have the identity
\bea
{\rm P } ( \varphi_j  | \theta_i  ) /{\rm P } ( \theta_i )  & = &   {\rm P }
( \theta_i  | \varphi_j  ) / {\rm P } ( \varphi_j ) \equiv {\rm P } (
\theta_i  \bigcap \varphi_j  )   \, . \label{j2}
\eea
in which the expression to the far right is known as the{ \it  joint
probability } for propositions $ \theta_i $  and $ \varphi_j $. The
discrepancy between (\ref{j1}) and (\ref{j2}) is an expression of the well
know fact
that joint probabilities, by definition symmetric, $  \theta_i  \bigcap
\varphi_j  = \varphi_j  \bigcap \theta_i     $, {\it do not exist}  for
mutually noncommuting experimental arrangements \footnote{ for a discussion
in terms of corresponding random variables see ref. \cite{43} }; the
experimental determination of one distribution fundamentally disturbs the
other \cite{3,6,16,17,42}, a case of which illustrated here Venn diagrams
\begin{center}
\scalebox{0.8}[0.8]{\includegraphics{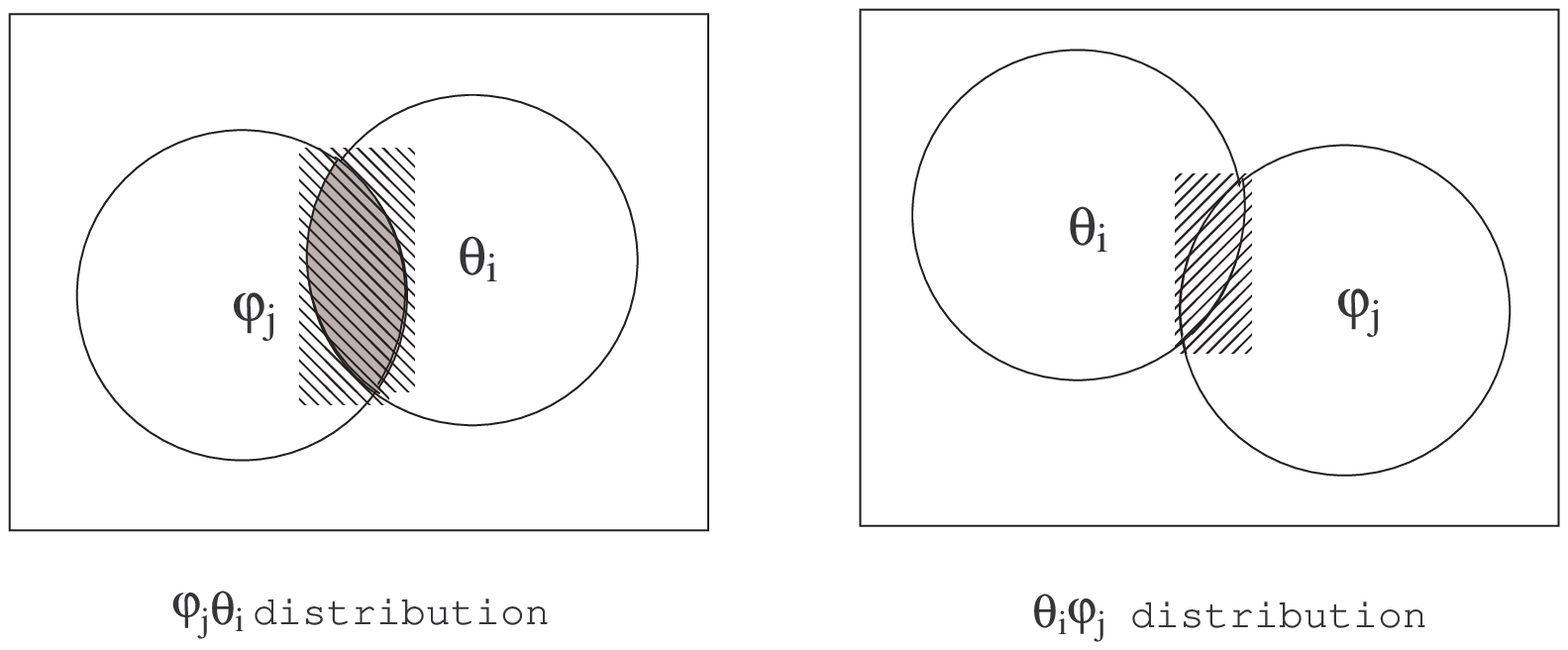}}\\
figure 6
\end{center}
so that $ \varphi_j \theta_i \neq \theta_i \varphi_j $ , propositions $
\theta_i $ and $ \varphi_j $  noncommuting.

\subsection{ nondistribution of probability}

There is thus an ambiguity when signifying the conditional probabilities of
noncommuting propositions in the usual set theoretic {\it notation} . The
notation is also standard to classical logic whose matrix of propositions,
like that of naive set theory, lacks the temporal ordering of events
needed to account for noncommutivity. Martin Strauss, a natural philosopher
who has written extensively on the subject of QM interpretation, long ago
made the point that, "Classical probability based on
the usual propositional calculus which is isomorphic to the set-theoretic
system of subsets of a given set has for its probability functions a domain
which is likewise isomorphic to the system, and {\it assumes} therefore
the simultaneous decidability of any two propositions...", the quote taken
from Jammer \cite{8}. In a more suitable though unfortunately less
common notation we have that $ {\rm P } ( \varphi_j  | \theta_i  ) /{\rm P }
( \theta_i ) \equiv {\rm P
} ( \theta_i ,  \varphi_j  )   $ \cite{29}, so that

\bea
{\rm P } ( \theta_i ,  \varphi_j  )    &=  &  {\rm P } ( \varphi_j ,
\theta_i  ) \equiv   {\rm P }(\theta_i  \bigcap \varphi_j^\prime) , \quad
\quad { \rm only \,\,in \,\, the \,\, event \,\, that \,\, } \theta_i
\varphi_j
  = \varphi_j  \theta_i   \, .\nonumber
\eea
On the other hand, from Eqn (\ref{marg}) we have the always valid identity
\bea
{\rm P } ( \theta_i ) &  = &  {\rm P } ( \varphi_j , \theta_i ) +   {\rm P
}( \varphi_j^\prime , \theta_i )  \nonumber
\eea
although, again, generally
\bea
{\rm P } ( \theta_i ) &\neq &  {\rm P } ( \theta_i  \bigcap \varphi_j ) +
{\rm P }(\theta_i  \bigcap \varphi_j^\prime)  \, .
\label{nd2}
\eea
Relation (\ref{nd2}) states a nondistribution, a violation of the usual distribution of
relative
probabilities. As the
noncommuting
conjunction itself, $ \theta_i \bigcap \varphi_j $, does not exist
as an experimental proposition \cite{6,17,42}, the nondistribution is
based ontologically.

\subsection{syntactic and semantic conjunction}

In the nondistributive
relation of the preceding section, Eqn. (\ref{nd1}), the conjunction appears
as a syntactic construct, an element in the object language of an X-space
generated sigma-algebra, whereas above in (\ref{nd2}) the conjunction corresponds to an
atomic event and
thus enjoys the status of a  primitive whose semantics derive entirely
within the meta-language of the probability theory \cite{8, 49}. As
classical probability spaces are induced by their corresponding experimental
outcome sets \cite{43}, the former syntactic conjunction belongs to one 
probability space (an $ H \bigcup H^\prime $ subspace), the
latter semantic conjunction to another (product of subspaces, $ H \times
H^\prime  $, itself however not a product space \cite{13, 30}). This unfortunate 
formal mis-identification has not helped resolve confusions in an already contentious
QM interpretations debate \cite{55}. Let us be especially clear on
the point we have just made: It is the syntactic nondistributivity of
propositions, equation (\ref{nd1}), that characterizes classical or
Boolean logic,
\begin{center}
syntactically distributive logic $ \Rightarrow $ classical logic
\end{center}
whereas the semantic nondistributivity of probabilistic measurements,
equation (\ref{nd2}), is characteristic of noncommutative or 'classical'
probability,
\begin{center}
semantically distributive probability  $ \Rightarrow $ classical
probability.
\end{center}
Whether the two relations express the same physics is a question taken up in
the next section.

\subsection{philosophical differences}

Several adherents to the conventional interpretation such as von Neumann,
Piron, and Jammer have offered specific examples of the physics behind the
noncommutative conjunction. Bohm, Bub and Bell in response maintain by
means of other semantics that such a conjunction does not exist, that
there can be no physical correspondence: To quote Bohm "...a and b,
represented by
noncommuting projection operators, can both be true with certainty if they
are confirmed as such by corresponding processes, whereas (because of
interference) {\it no process exists }to verify the proposition a $ \bigcap
$ b. In this case $ \omega( a \bigcap b ) = 0 $ without excluding the
possibility  $ \omega (a) = \omega(b) = 1 $ " \cite{6}. On the other hand,
it appears to be clear that upon the criteria set down by von Neumann and
others, the conditions, $
\omega (a) = \omega(b) = 1 $, have the very meaning, $ \omega( a \bigcap b
)
= 1 $ ... And then from Bell, "We are not dealing in B [a system of
experimental propositions] with logical
propositions, but with measurements involving for example differently
oriented magnets. The axiom [if $  \lan a \ran = \lan b \ran = 1 , \,\, {\rm
then \,\, } \lan a \bigcap b  \ran = 1  $ ] holds for quantum mechanical
states. But it is a quite peculiar property of them and in no way a
necessity of thought". Both criticisms appearing within the context of
their respective refutations of the logical no-go theorem are it seems
well-founded,
though they might have been better, more effectively, put within the context
of a larger, more
comprehensive assessment of the logical no-go
theorem, within the context of a kind of critical analysis that of the Bell's theorem, e.g.,
may be found everywhere \cite{11,19} \footnote{ Bohm might have added within this context
'that propositions cannot be decided upon simultaneously (as
with the incompatible semantic conjunction) does not impose that they are simultaneously undecidable (as in the case of the
incompatible  syntactic conjunction)', or Bell, that 'the impossibility of
simultaneous incompatible measurements (as in the case of the incompatible
semantic conjunction, $ \lan a \bigcap b  \ran = 0  $) does not preclude
the
simultaneous possibility of incompatible measurements (as in the case of the
incompatible syntactic conjunction $  \lan a \ran ,  \lan b \ran \neq 0 $)'
. And so forth.  As it happened, each man instead wage direct
assaults on the notion of realist semantic distributivity without shedding
much light on the seeming validity of syntactic distributivity, i.e. without
resolving the confusion between the two relations, a confusion that persists
to this day.
The same
may be said in respect of Strauss's 'complementary logic' \cite{17} and
Suppes's later version, both of which effectively  make the point though
without
an adequate appreciation for the distinction between compound and elementary
'events', syntactic and semantic conjunction....  and consequently
confounding
the notion of the 'classical' - of classical logic with classical physics.
This, from the syntactic rule of Strauss (long before the work of Bohm and
Bell) to Gudder's restriction to {\it experimental questions} \cite{16}
and several analysis that have appeared since. It comes as little surprise
that with few exceptions it is syntactic distributivity that holds
the interest of mathematics and philosophy (as evidenced in the works of
mathematicians and philosophers), while the empirically inclined
experimentalists have  concerned themselves primarily with the semantic, as
only the semantic noncommutative conjunction meets the criteria to describe
physical events (or non-events) \cite{37} }.

The logical no-go theorem mistakes syntactic distributivity for semantic
distributivity and thus confounds the realist notion of deterministically
possessed values with the notion of the commutation of incompatible
experiments.
While the two are certainly conceptually distinct, there remains the
possibility that the relations are empirically
related.

\section{ reconsideration of the argument }

Then what of the syntactic nondistributivity at the heart of the logical
no-go theorem? What does it signify, and what,
physically, would constitute an instance of it; how might such an instance
be confirmed or
refuted? The answer to these questions, as anticipated in the
previous section, supervenes on the precise
semantics associated with the noncommutative conjunction, for which, as
initial
guidance we may take the examples offered separately by
von Neumann, Piron and Jammer.

\subsection{syntactic and semantic physical distinction}

We consider an
experiment in which the incompatible observables, a and b, are randomly
sampled over an ensemble of 'identical' systems - an ensemble. Then,
according to our three authors, given a sufficiently
large sample we should never find that both P(a) = 0 or 1 and P(b) = 0 or 1.
I.e., we should never find that for each proposition, a and b, the
individual measurement outcomes
are all identical, either all yes, or all no. The equation, $ {\rm P } (a
\bigcap b ) = 0  $, thus states an instance of this, of the impossibility
of
finding the ensemble simultaneously in eigenstates of propositions a and b.
But clearly,
this is identical to the condition of dispersion itself, and furthermore
does in no obvious way
speak to the relevant question of measurement noncommutivity. Might
these simple points
have escaped their notice. Probably not. It is more likely that our authors
thought it reasonable to require of realist {\it outcome} states the same
statistics as QM eigenstates. We illustrate the assumption
for nondispersive state $ \psi $ with truth value 1 for
experimental proposition  $\theta_i $ ( and also $ \varphi_j $ by simply making the exchanges, $\theta \leftrightarrow \varphi $ and $i \leftrightarrow j $ )
\begin{center}
\scalebox{0.9}[0.9]{\includegraphics{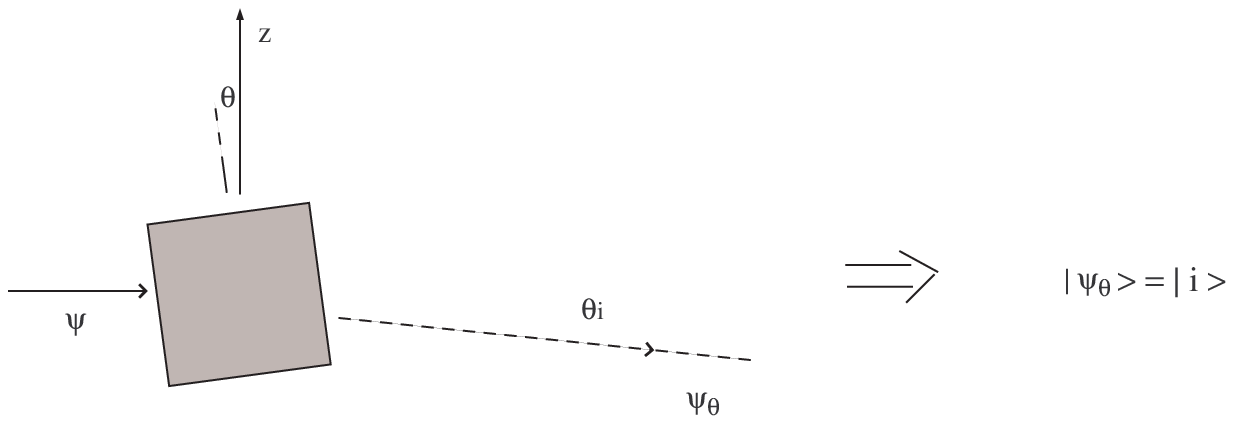}}\\
figure 7
\end{center}
Then,  $ P_j P_i | \psi \ran = P_j | i \ran $ and $  P_i P_j |
\psi \ran = P_i | j \ran  $, from which we have the probability relation,
P($\theta_i \varphi_j $) = P($ \varphi_j \theta_i $), same as for the
case of QM states under the operation of mutually commuting experiments.
There is however no reason a priori to require such statistics of
subensembles of
dispersion-free states, and it is easy to construct an empirically
consistent  measurement picture in
which they would not hold, as the
time evolution of a nondispersive state may depend, very reasonably,
upon the measurements performed on the representative system. Then by parameters
$
\lambda_{\theta \varphi } $  and $ \lambda_{ \varphi \theta }$ let us denote
this dependence in the case of two noncommutative temporal
sequences of experiments $ \theta $  and $ \varphi $ , and relate the state
of the system prior to measurement, $ \psi $(t), to its state following
measurement, $ \psi $(t'), by means of a time evolution operator U which
propagates the state vector from t to t' :   $ \psi(t) \rightarrow \psi
(t^\prime ) = U (\lambda_{\theta \varphi } ; t^\prime, t ) \psi (t) , \, \,
U
(\lambda_{ \varphi \theta } ; t^\prime, t ) \psi (t)  $. These then describe the
two distinct subensemble experimental processes
\begin{center}
\scalebox{0.9}[0.9]{\includegraphics{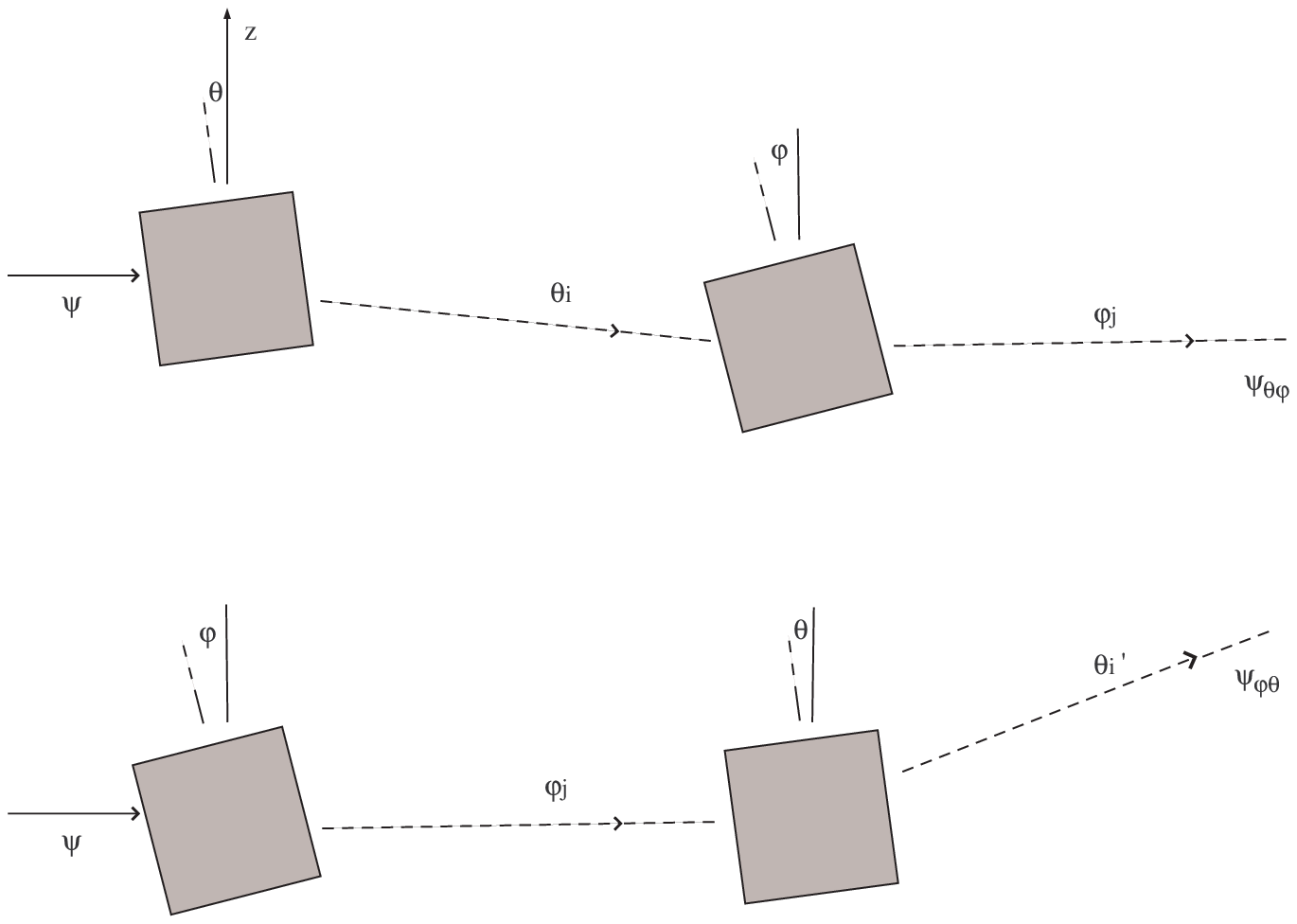}}\\
figure 8
\end{center}
So that while each initial subensemble is nondispersive, $
\omega_{1,2} (\theta_i ) , \, \omega_{1,2} (\varphi_j) = 1 \,\, {\rm or }
\,\, 0  $ (though with a combined non-zero dispersion:  $
\omega (\theta_i ) , \, \omega (\varphi_j) \neq 1 \,\, {\rm or } \,\, 0  $
), in violation of syntactic nondistributivity, eqn.(\ref{nd1}), the
consecutive measurement of incompatible propositions here maintain
noncommutivity, $ {\rm P }(\theta_i \varphi_j ) \neq  {\rm P }( \varphi_j
\theta_i )  $, thus satisfying the constraints of semantic
nondistributivity, eqn. (\ref{nd2}), the one condition thus
independent of the other.

In addition, this constraint linking nondistributivity to
noncommutivity and necessary to the logical no-go theorem is not amenable to
empirical testing, as the nondispersive ensembles themselves exist solely as
hypotheticals, within abstract partitions of the physical dispersive
ensembles of experience. To
again quote Bell \cite{3}, there is 'no necessity of thought' by which
dispersionless subensembles must obey
the statistics of QM, none then leading from nondispersion
to commutivity, a violation of one not infringing upon the possible
validity of the other. We may well accept then the existence of
syntactically
distributive, hence nondispersive, subensembles without
contravening their semantic nondistributivity, i.e. the noncommutivity of
incompatible measurements performed on all subensembles. This assertion, let
us
be clear, is at odds with the claim made by Jauch and Piron in
Ref.\cite{12} that 'The detailed analysis of this relation [syntactic
distributivity] shows that it has exactly the properties one would associate
with measurements which can be performed simultaneously without disturbing
each other [noncommutivity].', which continues, 'For instance, if the
propositions are represented by projection operators in a Hilbert space
..... '. As it turns out, the cited "instance" serves also as the sole
'detailed analysis' ever offered or referenced by the authors in
support
of the claim, and it seems indeede the only one available. Thus, the statistical
constraint implicit to the logical no-go theorem and imposed on
nondispersive states rests finally on the premise that all states are
quantum mechanical and hence dispersive. The argument is remarkable in
its blatant
circularity: All states dispersive $ \rightarrow $ No states nondispersive.
Unlike the hv's argument advanced by von Neumann, however, recognized only
a
full 30 years after publication as simple fallacy, this basic circularity in
the logical no-go argument did not long go unnoticed and was at publication quickly pointed out
(in another context) by Bohm \cite{6, 43} among others.

\subsection{formal refutation}

To continue one more, this problem with the logical no-go may also be
understood from a more formal standpoint, physical considerations aside. The law of distribution holds
whenever its components are sets, though not when they are spaces, as
formally defined. In fact, when we faithfully
respect component status the
nondistributivity inequality (\ref{nd1}) is immediately made distributive.
Generally, given any outcome space X $ = \bigcup_\alpha \Omega_\alpha =
\{ a, b, c, d, e, f, g, \ldots \} $ :
\bea
a &=& a \bigcap {\bf I } \nonumber \\
&=& a \bigcap ( b \bigcup b^\prime ) \nonumber  \\
&=& ( a \bigcap b ) \bigcup (a \bigcap b^\prime ) \nonumber \\
&=& ( a \bigcap b ) \bigcup  [  a \bigcap ( a \bigcup c  \bigcup d \bigcup e
\bigcup f \bigcup g \bigcup h \bigcup    \ldots )  ] \nonumber \\
&=& ( a \bigcap b ) \bigcup   a \bigcap ( a \bigcup c  \bigcup d \bigcup e
\bigcup f \bigcup g \bigcup h \bigcup     \ldots )  \nonumber \\
&=& a \bigcap [ (a \bigcap a^\prime   ) \bigcup (b \bigcap b^\prime   )
\bigcup(c \bigcap c^\prime   ) \bigcup(d \bigcap d^\prime   ) \bigcup \ldots
   ] \nonumber \\
&=& (a \bigcap a) \bigcup (a \bigcap a^\prime ) \bigcup (a \bigcap b)
\bigcup (a \bigcap b^\prime ) \bigcup (a \bigcap c) \bigcup (a \bigcap
c^\prime ) \bigcup (a \bigcap d) \bigcup (a \bigcap d^\prime ) \bigcup
\ldots  \nonumber \\
& =& a    \label{dis1}
\eea
where, (a $ \bigcap $ x) = 0 \quad whenever, a $ \neq $ x.

Syntactic nondistribution formally proceeds then from a set
theoretic inconsistency, where the compliment-defining universal set
\cite{31} on the rhs of (\ref{nd1}) is outcome set $ \Omega_b = \{b_1,  b_2, b_3, \ldots
, b_n  \}    $, but on the lhs is taken to be the outcome space X, thus yielding the
(\ref{nd1}) inequality
\bea
a \bigcap ( b \bigcup b_X^\prime ) & \neq & a \bigcap ( b \bigcup
b_\Omega^\prime )  = a \bigcap b + a \bigcap b^\prime_\Omega = 0 + 0 \quad
\quad \quad {\rm whenever} \, \, a \bigcap \Omega_b = 0 .     \nonumber
\eea
Physically, of course, $ b \bigcup b_X^\prime $ is not an experimental
proposition.

Alternatively, we may assign the inconsistency to the connective
conjunction itself, $ \bigcup
$, which joins two propositions on one side of (\ref{nd1}) to give their
(QM) span, while on the other their usual set theoretic (classical) union
\cite{62}.

In addition to the critical analysis of Bohm, the philosopher Popper was
also quick to weigh in,
pointing out, in a
lively exchange with the logical no-go authors an inconsistency in their
reasoning \cite{9}. Though certainly blatant (  $ ( b \bigcup
b^\prime ) \rightarrow ( b \bigcup b_X^\prime ) = ( b \bigcup b^\prime )   $
, nothing more, writes Pooper, than 'a simple slip') the misrepresentation
is obscured
by its logical idiom. Now at a safe distance one may appreciate the extent
to which this notational slip accommodates
the interpretation of the Copenhagen school: As $ \Omega
$ here designates the set of possible outcomes of an
experiment performed on an individual ensemble member, e.g. on an individual
particle, their truth values in the realist interpretation bivalent,
the outcome space X, on the other hand, as a union of such outcome sets, $
\bigcup_n \Omega_n   $, then designates the set of possible outcomes
corresponding to groups of
experiments performed upon groups of individual systems, upon ensembles, their truth values in any interpretation generally
non-bivalent. To substitute one for the other here would be to mistakenly impose a
constraint
existing within one interpretation, the orthodox interpretation, in the
course
of eveluating an opposing interpretation, the realist interpretation
\cite{43}... ; formal
sets always satisfy relation (\ref{nd1}), spaces generally do
not, and the very
possibility of a consistent realist interpretation is immediately ruled out
when it is assumed that H-space vectors represent all possible physical
states. And so, again, the circularity.

\section{ summary and conclusions }

We have shown that the logical no-go theorem involves the fallacy which
argues that the empirical validity of QM maintains only to
the exclusion of other descriptions, specifically, to the exclusion of a
presumably more complete realist hv theory, that because observed ensemble
states
are all quantum mechanical, so too must their constituent subensembles down to the single element be.
The theorem
assumes, in the words of Bohm, "that the current linguistic structure of QM
is the only one that can be used correctly to describe the empirical facts
underlying the theory" \cite{6}, but from Bell, "only QM averages over the
dispersion free states need produce this [observed statistical] property."

\subsection{differing views on probability theory}

If one accepts the proposition that qm states and observables refer
solely to
ensembles and their averages - a proposition accepted, in fact, by all
practicing
experimentalists as operationally valid - there opens the possibility of
the existence of a distinct
underlying individual ensemble-member reality, much as the Newtonian reality
of individual systems underlies the dynamics of statistical mechanics, as
the individual molecular reality of mutual electromagnetic interactions
underlies the ideal gas law, etc. Such is a {\it frequentist} understanding
of QM
probabilities as nothing more than the relative frequencies of ensemble
measurement outcomes. This is the view taken e.g. in Ballentine's
statistical or {\it ensemble} interpretation of QM \cite{15,32} where the
absence
of a exhaustive description of individual ensemble member reality casts QM
as a theory incomplete and possibly provisional, as an approximation to a
more
complete theory, again, much as statistical mechanics is incomplete with
respect to
Newtonian mechanics. This was also generally the view of QM statistics
held
by Einstein
\cite{32}.

From an opposing point of view, QM probabilities refer rather to
individual
events, to the outcomes of individual measurements made upon individual
ensemble members. In this case, empirical ensemble frequencies relate to the
'likelihoods' of particular individual outcomes. Thus, the non-bivalence
ordinarily characteristic of ensemble propositions are
immediately manifest in the propositions regarding measurements on
individual ensemble members.
And so, to refer back to the experiment of fig. 1, not only is the
spin projection of a particle not known before it has been measured
(epistemically non-existent), but at such a stage
the particle does not properly possess a spin (ontologically
non-existent);
it may only be said that the particle possesses a kind of likelihood that
a given projection outcome will be obtained upon measurement. Such is a
subjective {\it Bayesian} view of probability \cite{33,39} and is more or
less in
line with the orthodox interpretation of QM as a statistical
though physically complete theory.

When experiments share
a common outcome, i.e., when their outcome sets overlap, the theory
predicts and experiment verifies that the relative frequencies for those
outcomes are equivalent, independent of experimental context. This peculiar
{\it noncontextual} behavior of microscopic ensemble statistics thus becomes
by
interpretation characteristic also of individual microscopic measurements:
propositionally, $ \theta_i \bigcap \Omega_\theta = \varphi_j \bigcap
\Omega_\varphi  $ in the event that we have physically $ \theta_i = \varphi_j  $.

\subsection{noncontextual probability space X}

The conventional interpretation necessary to the logical no-go theorem is
effectively axiomatized by means of the probability space transformation
(\ref{space}), $ \{
\Omega_n \} \rightarrow   \bigcup_n \Omega_n  =  $  X . Earlier we spoke of this as a "generalization" of the probability theory, but the
term is not really appropriate here, as the loss of generality is indeed enormous; rather, the probability mapping is now {\it
specialized} to the case of noncontextual outcomes. The famous theorem of
Gleason \cite{7} illustrates the point:  Given the class of outcome sets $
\Omega_n $  of rank $ > $ 3, it is not possible while also
respecting the empirical sum rules, $ {\rm P } (\Omega_n ) = 1 $, i. e.,  $
{\rm P } (\Omega_n \bigcap {\rm X } ) = 1 $ for all $ \Omega_n $, to map the
corresponding
propositions to a bivalent state space, P : X $ \rightarrow $ \{0,1\}. It
thus
follows
from the harmless identity beginning our equation
(\ref{dis1})
\bea
a &=& a \bigcap ( b \bigcup b^\prime )          \nonumber
\eea
that
\bea
{\rm P }(a) &=& {\rm P }[ a \bigcap ( b \bigcup b^\prime )  ] \,  \quad
\quad \quad {\rm for \,\, all }\,\, a           \nonumber
\eea
constraining at least some of our propositions, a $ \in \Omega_a $, to now
non-bivalent truth values - a constraint, when
individual measurement outcomes are assumed independent of experimental
context, strictly enforced by Gleason's theorem. The probability space
transformation, this generalization, affects
also of course a corresponding shift in the logical form of the involved experimental
propositions, a shift from the properly conditional propositions of
experimental physics (with
experimental
contexts antecedently given or understood. E.g., a proposition $a_i$ which
says that an outcome $a_i$ will be obtained {\it when} or {\it if} an
$\Omega_a$ experiment is performed) to propositions that are
categorical. Thus from
propositions epistemic to those whose claims are ontological
\cite{34} \footnote{ Classically, this amounts to taking weight rather
than mass as an intrinsic property of a body. In practice, the intrinsic
masses are indeed determined, typically, from weight measurements, though
not
without the law of gravity and a consideration of experimental context (mass
of the planet on which the weight measurement is taken) }. When with this
noncontextuality (which, again, constrains a propositional equivalence  $
\theta_i =
\varphi_j $, whenever holds the physical outcome equivalence $ \theta_i =
\varphi_j $ )  is assumed in addition the principle of the value
definiteness, $ {\rm P }(\theta_i ) , {\rm P }(\varphi_j )  \in \{ 0, 1  \}
  $, which holds for realist dispersion-free states, we have that the
outcome
of a measurement $ \theta_i $ taken in the experimental context $ \theta $
  is the same as it would have been had the measurement been taken instead
in the experimental context $ \varphi $
\begin{center}
\scalebox{0.7}[0.7]{\includegraphics{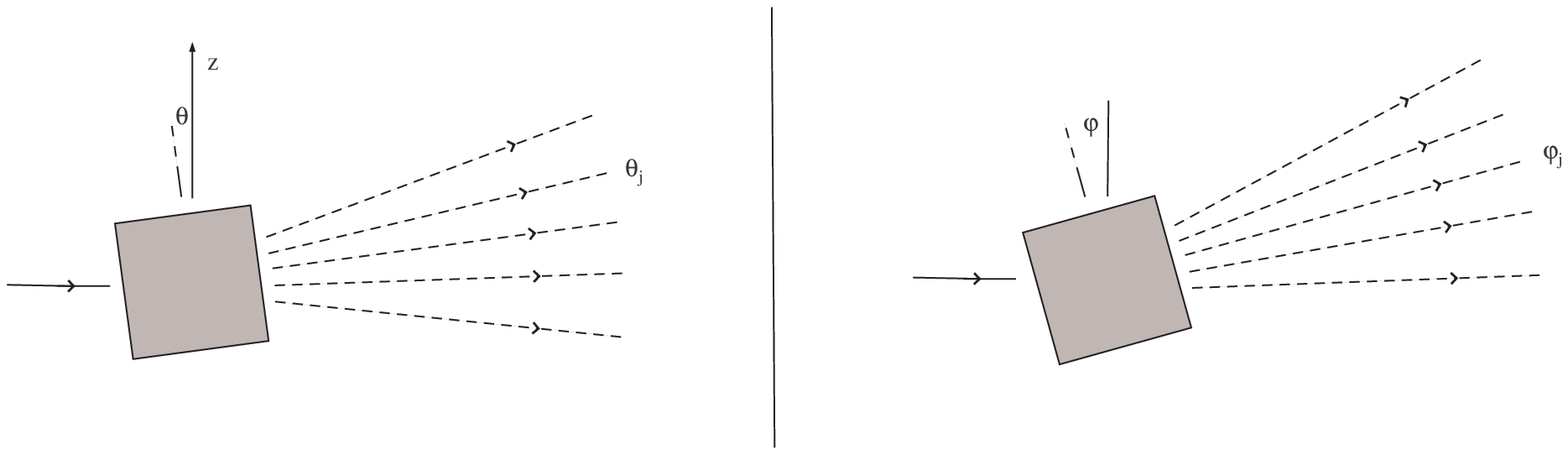}}\\
figure 9
\end{center}
a statement of the counterfactual definiteness of events, often
understood under appropriate conditions as a formulation of determinism
\cite{35, 20}.

This is also the constraint at the heart of the later and related no-go theorem of
Kochen and
Specker
\cite{4, 27, 20}. To briefly state, the KS analysis assumes as a
necessary element to any consistent realist view the noncontextual
embedding of QM observables in a nondispersive theory by means of the map,
P( $ \Omega_n \bigcap $ X ) = 1  for all $ \Omega_n $  (respecting sum
rules). But Gleason's already analytically rules out the possibility of such an embedding
\footnote{It is
remarkable that some, particularly mathematicians, seem at an utter loss why
anyone would want to imagine a {\it contextual} embedding, X  $
\rightarrow \{ \Omega_n \} $; physicists generally have less
trouble. Remarkable that anyone, including mathematicians \cite{36}, would
think a value-definite embedding of QM observables necessary to a
realist hv interpretation, a realist reading that {\it begins}, after all,
with
the set of physical outcome sets $ \{ \Omega_n \} $. But the thinking is
not
so uncommon as one might hope. At least not so in the opinion of S.
Goldstein \cite{44}: " In view of the radical character of quantum
philosophy, the arguments offered in  support of it have been surprisingly
weak. More remarkable still is the fact that it is not at all unusual, when
it comes to quantum philosophy, to find the very best physicists and
mathematicians making sharp emphatic claims, almost of a mathematical
character, that are trivially false and profoundly ignorant." An
exception are the views expressed by philosopher H. Stein who in 1970
writes,
"There is no obvious way to interpret eventualities [QM
projectors] as units of discourse, except under the conditions of a test
realizing them (when an eventuality may be correlated with the proposition
that {\it that eventuality holds}); but the eventualities realized
together in any experiment have ordinary Boolean logical relations to one
another, so that no non-standard logic here comes into play" \cite{39} }.
Interest in the KS analysis itself however persists to this day, and its
relevant issues are taken up by the writer in separate paper \cite{20}.

\subsection{further speculations}

In respect of the KS analysis, let us add here briefly one telling point
of reference.
While the system of micro-experimental propositions in the work of Jauch and
Piron
has a nondistributive, non-Boolean structure, and forms a complete
orthocomplemented
lattice, the QM logic of Kochen and Specker, on the other hand, is
structured partially
distributive, partially Boolean, and forms an orthocoherent orthoalgebra
\cite{2,4}. Consider with this the existence of several well-known deterministic
models of microscopic phenomena that also account for the
noncommutation of
incompatible observables (e.g., the famous model by Bohm \cite{47} or a
more recent one by Aerts \cite{48}), of the presence of an elephant in the
room, and it is obvious that
it is not a question of the nondispersion or the noncommutivity of
measurements that is at the heart of popular dissatisfaction with
"classical" readings of QM; here, the 'logical' approach to the question
of hv's misses the point. What is seminal to popular
anti-classical sentiment in the case of QM interpretations and in the
natural sciences generally is an
enduring suspicion of the scientific enterprise that
explains, predicts, and manipulates \cite{46,61} by virtue of the properly
'Classical' operational hypothesis according to which physical causes are by
nature
nomologically epistemic \cite{58} - i.e., lawlike - and consequently deterministic 
\footnote{ - a synonymy embedded in everyday usage: The official speaking on
behalf of BP oil company in a news report on a recent refinery
accident (24 March 05) assures the public that they will find out "why it
took place, what were the causes".}. This, even against the stringent
empiricism insisted upon by so many founders of the quantum theory.
Such as that of W. Heisenberg \cite{57}, who was happy to evoke 
this kind
argument in defense of a QM
completeness under threat \footnote{ What the student and non-specialist may
find alarming - what alarms the
writer - is the strength of {\it conviction} openly expressed in the QM
foundations literature by parties on
all sides of the interpretation issue.... In an otherwise excellent review,
A. Fine writes "It should be clear by now that I find the instrumentalist
interpretation of
the theory repugnant. Indeed I find it sad that the discredited
philosophical positivism of the 1930's, away from those doctrines the
behavioral sciences are finally being weaned, should find its last ditch
supporters among the middle generation of physicists..." \cite{40}. I am inclined to agree; sad
indeed...,  but such intensity is perhaps less surprising when one
considers the also uncharacteristic
and much publicized early foundations debate between Einstein and Bohr
\cite{8}. Ballentine, under fire, maintaining his usual composure and clarity
laments that "The entirely reasonable question, 'Are there
hidden-variable theories consistent with quantum theory, and if so, what are
their characteristics?,' has been unfortunately clouded by emotionalism. A
discussion of the historical and psychological origins of this attitude
would not be useful here. We shall only quote one example of an argument
which is in no way extreme (inglis, 1961, p.4), 'Quantum mechanics is so
broadly successful and convincing that the quest [for hv's] does
not seem hopeful.' The vacuous characteristic of this argument should be
apparent, for the success of quantum theory within its domain of definition
(i.e., the calculation of statistical distributions of events) has no
bearing on the existence of a broader theory (i.e., one which could predict
individual events.)". \cite{15}}, suggesting a certain determinism (excuse me, determination) on the 
part of the anti-realist, the anti-determinist camp. That this does indeed 
appear to be the
central
issue in dispute is rarely noted in the literature (See e.g. Ref.\cite{16}),
and on those occasions, as here, only in passing.

Of the three impossibility proofs we have mentioned only Bell's examines
the hv idea on the relevant question of determinism; it is also the simplest of the three. Increasingly it is this theorem like the man that is the most interesting and compelling. A definitive test of the
inequality however (notwithstanding a chorus
of claims to the contrary \cite{59}) has yet to be run \cite{20,68}.


\vspace{10mm}

\begin{thebibliography}{}


\bibitem{64}
S. Aranel, "Macroscopic", {\it Wikipedia, the free encyclopedia} (May
2005), URL = $ \lan $  http://en.wikipedia.org/wiki/Macroscopic $ \ran $.

\bibitem{54}
F. Laloe, "Do we really understand quantum mechanics? Strange correlations,
paradoxes, and theorems", {\it American Journal of Physics} {\bf 69},
655 (Jun 2001).

\bibitem{18}
J. von Neumann: {\it Mathematical Foundations of Quantum Mechanics},
Princeton: Princeton University Press (1955).

\bibitem{200}
S. Afshar, "Wavefunction collapse", {\it Wikipedia, the free encyclopedia}
(May 2005), URL = $ \lan $  http://en.wikipedia.org/wiki/Wavefunction\_collapse
$ \ran $.

\bibitem{65}
V. Easton and J. MaColl: dispersion, {\it Statistics Glossary} (Sept.
1997),
URL = $ \lan $
http://www.stats.gla.ac.uk/steps/glossary/presenting\_data.html\#
Disp $ \ran $.

\bibitem{1}
Hájek, Alan, "Interpretations of Probability", {\it The Stanford
Encyclopedia of Philosophy (Summer 2003 Edition)}, Edward N. Zalta (ed.),
URL = $ \lan $
http://plato.stanford.edu/archives/sum2003/entries/probability-interpret/ $
\ran $.

\bibitem{10}
K. R. Popper: {\it Quantum Theory and the Schism in Physics}, London:
Hutchinson \& Co. (1982).

\bibitem{11}
J. S. Bell: On the Einstein-Podolsky-Rosen paradox, {\it Physics} {\bf 1
}, 195 (1964).

\bibitem{19}
K. Williams, "A Question of Self-consistent Semifactuality" (Spring 2002),  
quant-ph/0512050.

\bibitem{20}
K. Williams, "Contextual Value-definiteness and the Kochen-Specker Paradox"
(Spring 2003), quant-ph/0512052.


\bibitem{100}
Eric W. Weisstein. "Bijection." From { \it MathWorld }--A Wolfram
Web Resource. URL = $ \lan $  http://mathworld.wolfram.com/Bijection.html $
\ran $.

\bibitem{110}
J. Wales, "Complementarity (physics)", {\it Wikipedia, the free
encyclopedia}
(May
2005), URL = $ \lan $
http://en.wikipedia.org/wiki/Complementarity\_\%28physics\%29 $ \ran $.

\bibitem{111}
S. Afshar, "Sharp complementary wave and particle behaviours in the same
welcher weg experiment" (May 2003), URL = $ \lan $
http://www.irims.org/quant-ph/030503/ $ \ran $; J. Wales, "Afshar
Experiment", {\it Wikipedia, the free encyclopedia}
(May 2005), URL = $ \lan $  http://en.wikipedia.org/wiki/Afshar\_experiment
$ \ran $.

\bibitem{12}
J. M. Jauch and C. Piron: Can Hidden Variables be Excluded in Quantum
Mechanics, {\it Helv. Phys. Acta} {\bf  36}, 827 (1963).

\bibitem{66}
G. Kimberling, "principle of bivalence", {\it Philosophy Pages}
(Oct.2004  ), URL = $ \lan $  http://www.philosophypages.com/dy/b2.htm\#biva
$ \ran $.

\bibitem{2}
Wilce, Alexander, "Quantum Logic and Probability Theory", {\it The Stanford
Encyclopedia of Philosophy (Spring 2003 Edition)}, Edward N. Zalta (ed.),
URL = $ \lan $
http://plato.stanford.edu/archives/spr2003/entries/qt-quantlog/ $ \ran $.

\bibitem{13}
J. M. Jauch: {\it Foundations of Quantum Mechanics}, London:
Addison-Wesley (1968).

\bibitem {14}
C. Piron: {\it Foundations of Quantum Physics}, London: W. A. Benjamin,
Inc. (1976).

\bibitem{26}
G. A. Gratzer: {\it Lattice theory; first concepts and distributive
lattices}, San Francisco: W. H. Freeman (1971).
P. Mittelstaedt: {\it Quantum logic}, Boston: D. Reidel Pub. Co., (c1978).

\bibitem{3}
J. S. Bell: On the Problem of Hidden Variables in Quantum Mechanics, {\it
Rev. Mod. Phys.} {\bf 38}, 447 (1966). Reprinted in {\it Speakable and
Unspeakable in Quantum Mechanics}, Cambridge: Cambridge  University Press
(1987).

\bibitem{6}
D. Bohm and J. Bub: A Refutation of the Proof by Jauch and Piron that Hidden
Variables Can Be Excluded in Quantum Mechanics, {\it  Rev. Mod. Phys.} {
\bf 38}, 470 (1966).

\bibitem{16}
S. P. Gudder: On Hidden-Variable Theories, {\it Rev. Mod. Phys.} { \bf 42
}, 229 (1968).

\bibitem{17}
M. Strauss: {\it Modern Physics and its Philosophy}  (Reidel, Dordrecht,
Holland, 1972), pp186-199.

\bibitem{42}
P. Suppes: The Probabilistic Argument for a Non-Classical Logic of Quantum
Mechanics, {\it Philosophy of Science} {\bf  33}, 14 (1966).

\bibitem{101}
V. Easton and J. MaColl: event, {\it Statistics Glossary} (Sept.
1997),
URL = $ \lan $
http://www.stats.gla.ac.uk/steps/glossary/probability.html\#event $ \ran $.

\bibitem{24}
G. Kimberling: positivism, {\it Philosophy Pages} ( Aug. 2002 ), URL = $
\lan $
http://www.philosophypages.com/dy/p7.htm\# posm $ \ran $.

\bibitem{25}
G. Kimberling: logical positivism, {\it Philosophy Pages} (Aug. 2002),
URL = $ \lan $  http://www.philosophypages.com/dy/l5.htm\# logp $ \ran $.

\bibitem{63}
Eric W. Weisstein. "Injection."From { \it MathWorld }--A Wolfram
Web Resource. URL = $ \lan $  http://mathworld.wolfram.com/Injection.html $
\ran $.

\bibitem{52}
C. Cohen-Tannoudji, B. Diu, and F. Laloe: {\it Quantum Mechanics}, New
York: John Wiley\&Sons (1977); M.D. Johnson, "Quantum Mechanics", {\it
Wikipedia, the free encyclopedia} (Feb. 2005), URL = $ \lan $
http://en.wikipedia.org/wiki/Quantum\_mechanics $ \ran $, and relevant
links.

\bibitem{56}
M. Hardy, "Stern-Gerlach experiment", {\it Wikipedia, the free encyclopedia
} (Sep. 2004), URL = $ \lan $  http://en.wikipedia.org/wiki/Stern-Gerlach\_
experiment $ \ran $.

\bibitem{38}
Eric W. Weisstein. "Probability Space." From {\it MathWorld}--A Wolfram
Web Resource. URL = $ \lan $
http://mathworld.wolfram.com/ProbabilitySpace.html $ \ran $,
and relevant links.

\bibitem{53}
E. Betts, "Hilbert Space", {\it Wikipedia, the free encyclopedia} (Dec.
2004), URL = $ \lan $  http://en.wikipedia.org/wiki/Hilbert\_space $ \ran $.

\bibitem{50}
G. Kimberling, " analytic/synthetic", {\it Philosophy Pages} ( Sep.2004
), URL = $ \lan $  http://www.philosophypages.com/dy/a4.htm\#ansy $ \ran $;
For a
different view, see, F. Damji, "W.V.Quine", {\it Wikipedia, the free
encyclopedia} (Jan. 2005), URL =$ \lan $
http://en.wikipedia.org/wiki/W\_V\_
O\_Quine $ \ran $.

\bibitem{45}
P. August, "Boolean Algebra", {\it Wikipedia, the free encyclopedia}
(Jan. 2005), URL = $ \lan $  http://en.wikipedia.org/wiki/Boolean\_algebra $
\ran $.

\bibitem{5}
G. Birkhoff and J. von Neumann: The Logic of Quantum Mechanics, {\it The
Annals of Mathematics} {\bf  37}, 823 (Oct. 1936).

\bibitem{8}
M. Jammer: {\it The Philosophy of Quantum Mechanics}, New York: John Wiley
\& Sons, Inc. (1974).

\bibitem{67}
J. Wales, "Law of excluded middle", {\it Wikipedia, the free encyclopedia}
(May
2005), URL = $ \lan $  http://en.wikipedia.org/wiki/Excluded\_middle $ \ran
$.


\bibitem{60}
Eric W. Weisstein: "Einstein-Podolsky-Rosen Paradox" From {\it Scienceworld}
-- A Wolfram Web Resource. URL = $ \lan $
http://scienceworld.wolfram.com/physics/Einstein-Podolsky-RosenParadox.html
$ \ran $;
Fine, Arthur, "The Einstein-Podolsky-Rosen Argument in Quantum Theory", {\it
The Stanford Encyclopedia of Philosophy (Summer 2004 Edition)}, Edward N.
Zalta (ed.), URL = $ \lan $
http://plato.stanford.edu/archives/sum2004/entries/qt-epr/ $ \ran $.

\bibitem{27}
Held, Carsten, "The Kochen-Specker Theorem", {\it The Stanford Encyclopedia
of Philosophy (Winter 2003 Edition)}, Edward N. Zalta (ed.), URL = $ \lan $
http://plato.stanford.edu/archives/win2003/entries/kochen-specker/ $ \ran $.

\bibitem{28}
Eric W. Weisstein, "Convex" From {\it MathWorld} -- A Wolfram Web
Resource. URL = $ \lan $  http://mathworld.wolfram.com/Convex.html $ \ran $.

\bibitem{41}
V. Easton and J. MaColl: probability, {\it Statistics Glossary} (Sept.
1997),
URL =  $ \lan $
http://www.stats.gla.ac.uk/steps/glossary/probability.html\#
Probability $ \ran $.

\bibitem{43}
A. Fine: Logic, Probability, and Quantum Theory, {\it Philosophy of Science}
{\bf  35}, 101 ( 1968).

\bibitem{29}
W. Heresiarch, "Conditional Probability", {\it Wikipedia, the free
encyclopedia} (Feb. 2004), URL = $ \lan $
http://en.wikipedia.org/wiki/Conditional
\_probability $ \ran $.

\bibitem{49}
S. Hocevar, "Metalanguage", {\it Wikipedia, the free encyclopedia} (Jan.
2005), URL = $ \lan $  http://en.wikipedia.org/wiki/Metalanguage $ \ran $.

\bibitem{30}
Eric W. Weisstein, "Direct Product" From {\it MathWorld} -- A Wolfram Web
Resource. URL = $ \lan $  http://mathworld.wolfram.com/DirectProduct.html $
\ran $.

\bibitem{55}
G. Kimberling, "Definition and Meaning", {\it Philosophy Pages} (Oct.2001),
URL = $ \lan $   http://www.philosophypages.com/lg/e05.htm $ \ran $.

\bibitem{37}
W. Heresiarch, "Event", {\it Wikipedia, the free encyclopedia} (Jan.
2005), URL = $ \lan $  http://en.wikipedia.org/wiki/Event $ \ran $.

\bibitem{31}
W. Heresiarch, "Universe (mathematics)", {\it Wikipedia, the free
encyclopedia} (Dec. 2004), URL = $ \lan $
http://en.wikipedia.org/wiki/Universe \_
\% 28mathematics \% 29 $ \ran $; W. Heresiarch, "Naive set theory", {\it
Wikipedia,
the free encyclopedia} (Dec. 2004), URL = $ \lan $
http://en.wikipedia.org/wiki/Na
\% EFve\_set\_theory $ \ran $.

\bibitem{62}
Geoffrey Hellman: "Quantum Logic and Meaning", {\it PSA: Proceedings of the
Biennial Meeting of the Philosophy of Science Association}, Volume Two:
Symposia and Invited Papers (1980), 493-511.

\bibitem{9}
K. R. Popper: Birkhoff and von Neumann's Interpretation of Quantum
Mechanics, {\it Nature} {\bf 219}, 682 (1968).

\bibitem{15}
L. E. Ballentine: The Statistical Interpretation of Quantum Mechanics, {\it
Rev. Mod. Phys.} {\bf  42}, 358 (1970).

\bibitem{32}
K. Aylward, {\it Quantum Mechanics. The Ensemble Interpretation} (Sept.
2004), URL = $ \lan $  http://www.anasoft.co.uk/quantummechanics/ $ \ran $.

\bibitem{33}
W. Heresiarch, "Bayesian Probability", {\it Wikipedia, the free
encyclopedia} (Jan. 2005), URL = $ \lan $
http://en.wikipedia.org/wiki/Bayesian\_
probability $ \ran $; Talbott, William, "Bayesian Epistemology", {\it The
Stanford
Encyclopedia of Philosophy (Fall 2001 Edition)}, Edward N. Zalta (ed.), URL
= $ \lan $
http://plato.stanford.edu/archives/fall2001/entries/epistemology-bayesian/ $
\ran $.

\bibitem{39}
H. Stein: Is there a Problem of Interpreting Quantum Mechanics?, {\it Nous}
{\bf 4}, 93 (1970).

\bibitem{7}
A. M. Gleason: Measures on the Closed Subspaces of a Hilbert Space, {\it J.
Math. \& Mech.} {\bf 6}, 885 (1957).

\bibitem{34}
Edgington, Dorothy, "Conditionals", {\it The Stanford Encyclopedia of
Philosophy (Fall 2001 Edition)},  Edward N. Zalta (ed.), URL = $ \lan $
http://plato.stanford.edu/archives/fall2001/entries/conditionals/ $ \ran $.

\bibitem{35}
I. Kvart: {\it A Theory of Counterfactuals},  Indianapolis: Hackett
Publishing Complny, Inc. (1985); G. Kimberling, "counterfactual", {\it
Philosophy Pages} (Aug.2002), URL = $ \lan $
http://www.philosophypages.com/dy/c9.htm\#couf $ \ran $; Menzies, Peter,
"Counterfactual Theories of Causation", {\it The Stanford Encyclopedia of
Philosophy (Spring 2001 Edition)}, Edward N. Zalta (ed.), URL = $ \lan $
http://plato.stanford.edu/archives/spr2001/entries/causation-counterfactual/
$ \ran $.

\bibitem{4}
S. Kochen and E. Specker: The Problem of Hidden Variables in Quantum
Mechanics, {\it J. Math. \& Mech.} {\bf 17}, 59 (1967).

\bibitem{36}
I. Pitowsky: Betting on the outcomes of measurements: a Bayesian theory of
quantum probability, {\it Stud. Hist. Philos. Sci.} {\bf  34} , 395
(2003).

\bibitem{44}
S. Goldstein: Quantum Philosophy: The Flight from Reason in Science, {\it
Annals of the New York Academy of Sciences} {\bf 775}, 119 (1996).

\bibitem{47}
D. Bohm, "A Suggested Interpretation of the Quantum Theory in Terms of
'Hidden' Variables, I and II", {\it Phys. Rev.} {\bf  85}, 166 and 180
(1952); K. Berndl, M. Daumer, D. Durr, S. Goldstein, and N. Zanghi, "A
Survey of Bohmian Mechanics", {\it Nuovo Cim.} {\bf B110}, 737
(1995). URL = $ \lan $  http://xxx.lanl.gov/abs/quant-ph/9504010 $ \ran $.

\bibitem{48}
D. Aerts: Quantum Structures - an attempt to explain the origin of their
appearance in nature, {\it Int. J. Phys.} {\bf 34}, 1165 (Aug.1995).


\bibitem{46}
S. Hocevar, "Philosophy of Science", {\it Wikipedia, the free encyclopedia}
(Jan. 2005), URL = $ \lan $
http://en.wikipedia.org/wiki/Philosophy\_of\_science $ \ran $;
G. Kimberling, "Scientific Explanation", {\it Philosophy Pages} (Sep.2004),
URL = $ \lan $  http://www.philosophypages.com/lg/e15.htm $ \ran $.

\bibitem{61}
A. Reingold, "Free will", {\it Wikipedia, the free encyclopedia} (Feb.
2005), URL = $ \lan $  http://en.wikipedia.org/wiki/Free\_will $ \ran $.

\bibitem{58}
Woodward, James, "Scientific Explanation", {\it The Stanford Encyclopedia of
Philosophy (Summer 2003 Edition)}, Edward N. Zalta (ed.), URL = $ \lan $
http://plato.stanford.edu/archives/sum2003/entries/scientific-explanation/ $
\ran $.


\bibitem{57}
W. Heisenberg: Über den anschaulichen Inhalt der quantentheoretischen
Kinematik und Mechanik, {\it Zeitschrift für Physik} {\bf  43}172 (1927).
For a recent account see: Krips, Henry, "Measurement in Quantum Theory",
{\it The Stanford Encyclopedia of Philosophy (Winter 1999 Edition)} , Edward
N. Zalta (ed.), URL = $ \lan $
http://plato.stanford.edu/archives/win1999/entries/qt-measurement/ $ \ran $.

\bibitem{40}
A. Fine: On the completeness of quantum theory, {\it Synthese} {\bf 29},
257 (1974).

\bibitem{59}
C. Thompson, "Bell test experiments", {\it Wikipedia, the free encyclopedia}
(Mar. 2005), URL = $ \lan $
http://en.wikipedia.org/wiki/Bell\_test\_experiments
\# Experimental\_assumptions $ \ran $; A. Whitaker, "John Bell and the most
profound
discovery of science", {\it Physics World} (Dec. 1998), posted at {\it
Physicsweb}, URL =  $ \lan $  http://physicsweb.org/articles/world/11/12/8/1
$ \ran $.

\bibitem{68}
C. Thompson, "Letter to AJP", (Aug. 2001), URL = $ \lan $
http://freespace.virgin.net/ch.thompson1/Letters/Laloe.htm $ \ran $.






\end{thebibliography}
\end{document}